\documentclass[twocolumn, prx, showpacs, amsmath, amssymb, superscriptaddress]{revtex4-1}

\usepackage{color}
\usepackage{soul}
\usepackage{bm}
\usepackage{bbm}
\usepackage{graphicx}
\usepackage{amsmath}
\usepackage{hyperref}
\usepackage{todonotes}
\usepackage{epsfig}

\def \be{\begin{equation}}
\def \ee{\end{equation}}

\begin{document}
\title{
Non-Abelian symmetries and disorder: a broad non-ergodic regime and anomalous thermalization\\
}

\author{Ivan V. Protopopov}
\affiliation{Department of Theoretical Physics, University of Geneva, 1211 Geneva, Switzerland}
\affiliation{L.\ D.\ Landau Institute for Theoretical Physics RAS, 119334 Moscow, Russia}
\author{Rajat K. Panda}
\affiliation{The Abdus Salam ICTP, Strada Costiera 11, 34151 Trieste, Italy}
\affiliation{Scuola Internazionale di Studi Superiori Avanzati, Via Bonomea, 265, 34136 Trieste, Italy}
\author{Tommaso Parolini}
\affiliation{Scuola Internazionale di Studi Superiori Avanzati, Via Bonomea, 265, 34136 Trieste, Italy}
\affiliation{INFN, Sezione di Trieste, Via Valerio 2, 34127 Trieste, Italy}
\author{Antonello Scardicchio }
\affiliation{The Abdus Salam ICTP, Strada Costiera 11, 34151 Trieste, Italy}
\affiliation{INFN, Sezione di Trieste, Via Valerio 2, 34127 Trieste, Italy}
\author{Eugene Demler}
\affiliation{Department of Physics, Harvard University, Cambridge, MA 02138, USA}
\author{Dmitry A. Abanin}
\affiliation{Department of Theoretical Physics, University of Geneva, 1211 Geneva, Switzerland}

\begin{abstract}

Previous studies revealed a crucial effect of symmetries on the properties of a single particle moving in a disorder potential. More recently, a phenomenon of many-body localization (MBL) has been attracting much theoretical and experimental interest. MBL systems are characterized by the emergence of quasi-local integrals of motion, and by the area-law entanglement entropy scaling of its eigenstates. In this paper, we investigate the effect of a non-Abelian $SU(2)$ symmetry on the dynamical properties of a disordered Heisenberg chain. While $SU(2)$ symmetry is inconsistent with the conventional MBL, a new non-ergodic regime is possible. In this regime, the eigenstates exhibit faster than area-law, but still a strongly sub-thermal scaling of entanglement entropy. Using extensive exact diagonalization simulations, we establish that this non-ergodic regime is indeed realized in the strongly disordered Heisenberg chains. We use real-space renormalization group (RSRG) to construct tree-tensor-network approximation to excited eigenstates, and demonstrate the accuracy of this procedure for systems of size up to $L=26$. As the effective disorder strength is decreased, a crossover to the thermalizing phase occurs. 
To establish the ultimate fate of the non-ergodic regime in the thermodynamic limit, we develop a novel approach for describing many-body processes that are usually neglected by RSRG. This approach is capable of describing systems of size $L\gtrsim 2000$. We characterize the resonances that arise due to such processes, finding that they involve an ever growing number of spins as the system size is increased. Crucially, the probability of finding resonances grows with the system's size. Even at strong disorder, we can identify a large lengthscale beyond which resonances proliferate. Presumably, this eventually would drive the system to a thermalizing phase. However, the extremely long thermalization time scales indicate that a broad non-ergodic regime will be observable experimentally. Our study demonstrates that, similar to the case of single-particle localization, symmetries control dynamical properties of disordered, many-body systems. The approach introduced here provides a versatile tool for describing a broad range of disordered many-body systems, well beyond sizes accessible in previous studies. 

\end{abstract}
\date{\today}

\maketitle

\section{Introduction}
\label{Sec:Introduction}

The remarkable experimental advances of the past decade have opened a window into probing highly non-equilibrium dynamics of interacting quantum systems, using platforms such as ultracold atoms~\cite{BlochColdAtoms}, trapped ions~\cite{Blatt12}, and NV-centers in diamond~\cite{Choi16DTC}. One fascinating outcome of this research direction was the discovery that strong quenched disorder can suppress thermalization in isolated, many-body systems. This phenomenon, termed many-body localization (MBL), has attracted a lot of attention, both theoretically \cite{Basko06,Mirlin05,OganesyanHuse,Znidaric08,PalHuse,Vosk13,Serbyn13-2,Serbyn13-1,Huse13,Moore12,ScardicchioLIOM,Alet14,Demler14,Ponte15,Lazarides15,Abanin20161,Khemani16} and experimentally \cite{Bloch15, Bloch16-2, Monroe16, xu2018, Shahar, Choi16DTC, Bordia17, Lukin2018} (see Ref.~[\onlinecite{AbaninReview}] for a recent review). MBL systems constitute a novel dynamical phase of matter, in which quantum coherence is long-lived and largely protected~\cite{Serbyn_14_Deer,Bahri:2015aa}. The fact that such systems break ergodicity and thus are not described by conventional statistical mechanics opens many new opportunities for quantum dynamics and, in particular, enables non-equilibrium phases in periodically driven systems~\cite{Ponte15,Lazarides15,Abanin20161,Khemani16,Else16}.

Much of the progress in describing MBL and related phenomena was driven by the realization that fully MBL systems of e.g. spins or fermions on a lattice exhibit a new kind of robust emergent integrability~\cite{Serbyn13-2,Serbyn13-1,Vosk13,Huse13,ScardicchioLIOM}. Specifically, it is a complete set of quasi-local integrals of motion (LIOMs) that underlies the ergodicity breaking in MBL phases.  The LIOM construction naturally explains the area-law entanglement of the MBL eigenstates~\cite{Serbyn13-1,Bauer13}, logarithmic entanglement growth in a quantum quench experiment~\cite{Znidaric08,Moore12,Serbyn13-1,Huse13,Lukin2018}  and a number of other dynamical properties of the MBL phase~\cite{Serbyn14,Vasseur14,Fischer}.

\begin{figure}[t]
\includegraphics[width=0.95\columnwidth]{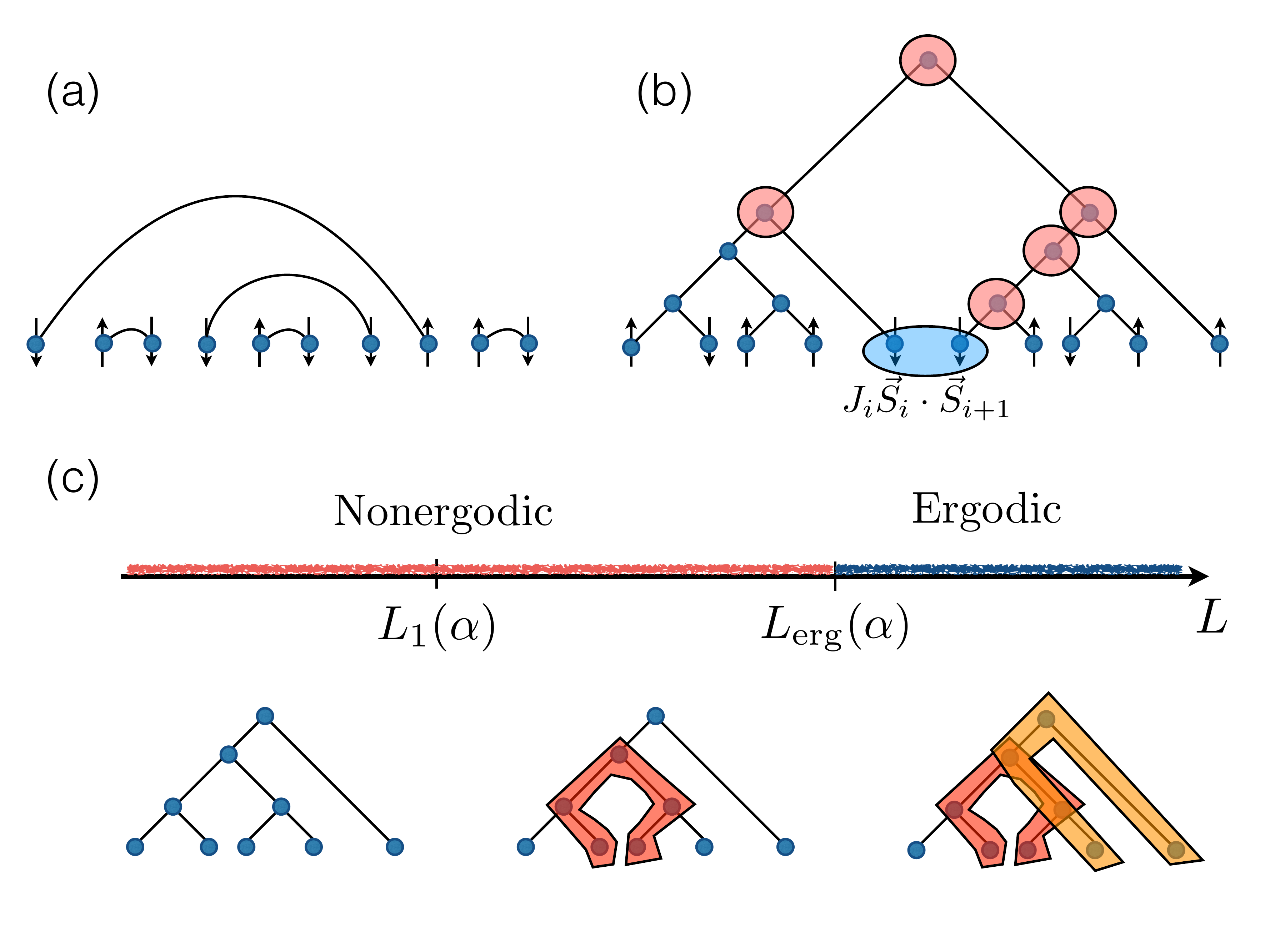}
\caption{(a) A cartoon of the ground state of a random antiferromagnetic Heisenberg chain; (b) Strong-disorder renormalization group aims to construct approximate eigenstates. It yields a tree state, characterized by its geometry and the choice of total block spins at each node (see main text). The Heisenberg Hamiltonian, written in this basis, gives rise to processes which can change the spins along the ``causal" path connecting two neighboring spins, to one of the block spins; (c) A schematic dynamical phase diagram of the random Heisenberg model. There are three regimes: (I) at short length scales, $L<L_1(\alpha)$, the SDRG tree states are accurate approximations to the eigenstates; (II) at intermediate length scales, $L_1(\alpha)<L<L_{\rm erg}(\alpha)$, there are resonances but the system remains non-ergodic; (III) above some large lengthscale $L>L_{\rm erg}(\alpha)$, the resonances proliferate and the system becomes thermalizing (see Sections \ref{Sec:Resonances_2} and \ref{Sec:Resonances_3} for a definition of these scales). }
\label{Fig:schematic}
\end{figure}

It quickly became clear  that distinct MBL phases are possible. Much like in the theory of thermodynamic phase transitions, the symmetry of the system plays a central role. For example,  systems with a discrete $\mathbb{Z}_2$ symmetry~\cite{Pekker14, Kjall14, Huse13L} can exhibit two distinct MBL phases: in one of them, the eigenstates spontaneously break the  $\mathbb{Z}_2$ symmetry, and in the other the symmetry is preserved. Both phases can be described using LIOM theory. 

Disordered systems with {\it continuous non-Abelian symmetries}, which constitute a broad and experimentally relevant class, show a qualitatively different behavior. An example of such a system is a disordered, $SU(2)$-symmetric spin chain. Crucially, a non-Abelian symmetry such as $SU(2)$ is inconsistent with all eigenstates obeying area-law entanglement entropy. Thus, conventional MBL with a complete set of LIOMs is forbidden by symmetry in this case~\cite{PotterSymmetry,Protopopov2017}. Some integrals of motion must become nonlocal; accordingly, the entanglement entropy of a subsystem in a typical, highly excited eigenstate must scale at least logarithmically with that subsystem size $\ell$ in 1D systems, $S_{\rm ent} (\ell) \gtrsim c\log(\ell)$ (where $c$ is a coefficient of order one). 

The fact that an $SU(2)$ symmetry enforces a minimum amount of entanglement in the eigenstates raises several fundamental questions. What is the nature of the excited eigenstates and the corresponding dynamical properties of disordered, $SU(2)$-symmetric systems? One exciting possibility hypothesized in Refs.~\cite{Agarwal2015,Protopopov2017} is that at sufficiently strong disorder a new kind of dynamical, non-ergodic phase may emerge -- characterized by the entanglement entropy of excited eigenstates that is sub-thermal, but scales faster than the area-law (e.g.\ as $S_{\rm ent} (\ell) \sim c\log(\ell)$). Such a phase would display only a partial set of LIOMs, being distinct from the conventional MBL phase.  Another, equally intriguing possibility is that thermalization may be inevitably enforced by such symmetries in thermodynamic limit~\cite{VasseurHotChains}. If this is the case, it would be highly desirable to understand the microscopic processes that govern thermalization, as well as the corresponding time- and lengthscales. 

This topic has been attracting strong interest, and several works provided valuable complementary insights into the above questions. Ref.~\cite{VasseurHotChains} have studied random $SU(2)_k$ anyonic chains, arguing that the breakdown of strong-disorder, real-space renormalization group (SDRG) approach as $k\to \infty$ signals self-thermalization of $SU(2)$-symmetric spin chains. Ref.~\cite{Agarwal2015} computed the noise spectrum of random Heisenberg chains using SDRG approach applied to excited states. Ref.~\cite{Protopopov2017} introduced a toy model, in which eigenstates of an $SU(2)$-symmetric spin chain are described by regular tree tensor networks with $S_{\rm ent} (\ell) \gtrsim c\log(\ell)$ entanglement entropy scaling; they studied the stability of such eigenstates under local perturbations of the Hamiltonian, finding indications of eventual slow delocalization. Further, Refs.~\cite{PrelovsekHubbard1,PrelovsekHubbard2} and Ref.~\cite{Bonca17} considered spin dynamics in disordered Hubbard and $t-J$ models, respectively. They found that spins were not localized even at strong disorder, and numerically studied spin transport, finding indications of sub-diffusive behavior. While transport does not imply ergodicity, this is another signal that in the presence of non-Abelian symmetries, a localized phase cannot have plain vanilla MBL phenomenology.

In this paper, we study $SU(2)$-symmetric disordered spin chains, focusing on their spectral properties, and the properties of highly excited eigenstates. The starting point of our analysis is the SDRG which is used to approximately construct excited eigenstates. This procedure, originally introduced to describe low-energy properties of random spin chains~\cite{ma1979random,Dasgupta1980,Fisher1992}, has been recently applied to the {\it highly excited eigenstates} in a range of systems~\cite{Vosk13, Pekker14,VasseurHotChains,Agarwal2015}. Applied to the random Heisenberg chains, SDRG yields a caricature of an eigenstate in the form of an (irregular) tree tensor network, with the structure that depends on the disorder realization; at each step of this construction, two spins which are strongly interacting with each other (relative to their interactions with their other neighbors) are added to form some other total spin. This is illustrated in Fig.~\ref{Fig:schematic}. Naturally, such states are strongly non-ergodic, although distinct from the conventional MBL eigenstates, e.g.\ in their entanglement properties (see below). So if the SDRG procedure remains accurate, the system is in a novel non-ergodic phase. However, typically the SDRG procedure only allows one to test for "local" resonances involving a small number of nearby spins, and therefore it is an open question when/whether SDRG is reliable and gives a good approximation to system's eigenstates at large system sizes. 

Below we investigate how well the SDRG procedure approximates system's eigenstates. To that end, we first perform extensive numerical simulations of spectral statistics and system's eigenstates. In particular, we will test the eigenstate thermalization hypothesis (ETH), which is believed to underlie thermalization in ergodic systems~\cite{Polkovnikov-rev}. We find that at strong disorder, there is a broad non-ergodic regime in which SDRG accurately captures the eigenstates. At weak disorder, above certain lengthscale, we find evidence for thermalization and breakdown of SDRG. We investigate how this lengthscale depends on the strength of the effective disorder, in the regime where it is smaller than the largest system size accessible numerically ($L=26$). 

To describe the behavior of large chains, far beyond those accessible via conventional numerical techniques, we develop a novel approach to describe nonlocal, multi-spin processes that are not captured by the conventional SDRG. For that purpose, we analyze the relevance of terms in the Hamiltonian that are responsible for the processes that are usually neglected in SDRG. These terms mix different states in the SDRG and if the mixing is sufficiently strong, they cannot be neglected and give rise to \emph{resonances}. We study how the number of resonances grows with the system's size, and describe their properties, such as energy scales and the number of physical spins involved. 

We find that at strong disorder the resonances are absent in a surprisingly broad range of lengthscales, signalling a regime in which SDRG describes eigenstates accurately.
Eventually, in sufficiently large systems, resonances will proliferate, perturbation theory in the terms neglected by the SDRG will not converge, and the system will thermalize. We expect this to give rise to full ergodicity, in an unconventional way that we will describe. Thus, our conclusion favours the scenario of "non-Abelian-symmetry-protected thermalization". Our work shows that this thermalization proceeds via long-range resonances that involve many spins; we extract the corresponding time scales, and find them to be extremely long at strong disorder. Thus, for all practical purposes, the strongly disordered system would appear non-ergodic, for reasonably short experimental observation times. 

The rest of the paper is organized as follows:
In Section II we introduce the model and describe the SDRG procedure which will be used to find approximate eigenstates. In Section III we first use exact diagonalization (ED) and a measure of participation ratios to check  how well the approximate eigenstates given by SDRG agree with the exact one. Then we investigate the onset of ETH and its breakdown at strong disorder using various measures (level statistics, statistics of matrix elements, and entanglement entropy). In Section IV we develop our SDRG-based approach to the analysis of resonances. We show how the terms neglected in the SDRG give rise to resonances which eventually proliferate, leading to thermalization of very large systems. Section V closes the paper with a recapitulation and suggestions for future work.

\section{Strong disorder renormalization group and tree states}
\label{Sec:RGandTreeStates}

We start this Section by introducing the model of a disordered  Heisenberg chain in Subsection~\ref{Sec:RGandTreeStates_1}. We refresh the well-studied example of a random-field Heisenberg model which lacks the $SU(2)$ symmetry, and compare it to the symmetric Heisenberg chain. We then review  the construction of the approximate eigenstates based on SDRG~\cite{ma1979random, Dasgupta1980, Fisher1992} paradigm. Basic properties of tree states obtained by SDRG are discussed (Subsection~\ref{Sec:RGandTreeStates_2}). Finally, in Subsection~\ref{Sec:RGandTreeStates_3} we qualitatively describe our approach to probing the stability of tree states obtained by SDRG. The detailed numerical studies are presented in the subsequent sections. 

\subsection{The model and preliminary remarks}
\label{Sec:RGandTreeStates_1}
The model we study  is the disordered, Heisenberg spin-$1/2$ chain with the Hamiltonian 
\begin{equation}
H=\sum_{i=1}^L J_i {\bf S}_i\cdot {\bf S}_{i+1}
\label{Eq:Ham}
\end{equation}
where ${\bf S}_i=(S_i^x,S_i^y,S_i^z)$ are the standard spin operators, and the couplings $J_i$ are independent random variables with a probability distribution $P(J)$ specified by two parameters, $\eta$ and $\alpha$. The parameter $0<\eta<1$ gives  the fraction of antiferromagnetic (positive) couplings in the system. Throughout this work we assume $\eta=0.5$. 
We do not expect the properties of highly excited eigenstates in the middle of the many-body band to exhibit a significant dependence on the choice of $\eta$. Note that the ground state properties do not depend strongly on the value of $\eta$ (except for the extremal points $\eta=0,1$ -- see \cite{Westerberg1997}). 

The  parameter $\alpha>0$ controls the distribution of $|J|$. We assume that the p.d.f. of this distribution has a  power-law form with a cutoff at $|J|=1$:  
\begin{equation}
P(|J|)=\frac{\alpha\Theta\left(1-|J|\right)}{|J|^{1-\alpha}},
\label{Eq:PJ}
\end{equation}
where $\Theta(x)$ stands for the Heaviside function. This distribution of couplings emerges naturally in a wide range of low temperatures, as it was shown in the seminal papers \cite{ma1979random, Dasgupta1980}. In that context, under the assumptions of what now would be called ETH, it can explain the anomalous exponent of the specific heat observed in early experiments \cite{PhysRevB.12.356}.

The exponent $\alpha$ effectively controls the strength of disorder, with smaller $\alpha$ corresponding to stronger disorder. Indeed, for the distribution (\ref{Eq:PJ}) the ratio of two neighbouring couplings in the system has a typical value
\begin{equation}
\left.\frac{\max(|J_1|, |J_2|)}{\min(|J_1|, |J_2|)}\right|_{\rm typ}\equiv \exp\left(\langle\left|\ln |J_1|/|J_2|\right| \rangle\right)=e^{1/\alpha}.
\label{Eq:disorder}
\end{equation}
This ratio increases exponentially when $\alpha\to 0$. Therefore, at small $\alpha$ it becomes more and more likely to find  exchange constants in the system that are {\it much} larger than the two neighbouring ones. This is exactly the condition that enables SDRG, as we discuss below. 

Another quantity of interest is the smallest coupling $J$ (in absolute value) in the whole system, representing the ``weakest link". We find that 
\begin{equation}
\min_i J_i\sim \alpha^{-1}\Gamma(1/\alpha) L^{-1/\alpha}.
\label{Jmin}
\end{equation}
For $\alpha=0.3$ and $L\simeq 20$ this coupling can be as small as $10^{-3} \langle J \rangle$ (here $ \langle J \rangle$ is the mean value of $J_i$).

Throughout the paper it will be helpful to contrast our findings to the properties of the random-field XXZ model, which has been studied extensively in the literature~(see  Refs.~\cite{ALET2018498, AbaninReview, ParameswaranVasseurReview2018}  for recent reviews):
 \begin{equation}
 H_{XXZ}=t\sum_{i}\left(S_{i}^{x}S_{i+1}^x+ S_{i}^{y}S_{i+1}^y\right)+U\sum_{i}S_{i}^zS_{i+1}^z +\sum_{i}h_i S_i^z
 \label{Eq:HXXZ}
 \end{equation}
 The model (\ref{Eq:HXXZ}) can be mapped, via Jordan--Wigner transform, onto an interacting fermionic problem with $t$ representing the hopping amplitude, $U$ the nearest-neighbour interaction, and $h_i$ the random on-site potential with a variance that we denote by $W$. 
 In the following, for concreteness, we will assume that $t\sim U$, such that disorder strength is described by a single dimensionless parameter $L_W=W/t$. Then, the XXZ model is known to have a diffusive--subdiffusive dynamical transition  \cite{vznidarivc2016diffusive} at $L_W\simeq 0.55$ and an MBL--thermal transition at $L_W\simeq 3.5$ \cite{Alet14}. 
 
 Note that in the limit of strong disorder, $W\gg t$, the parameter $L_W$ can be interpreted as a typical distance between (rare) pairs of ``resonant'' sites in the model that happen to have close enough values of the magnetic field to enable resonant spin exchange (or, equivalently, hopping in the fermionic model). A resonance between spins 1 and 2 appears, e.g. if $|h_1-h_2|\lesssim t$.  Starting at a very large disorder these resonant sites are typically well separated by distances of $O(L_W)$, and one can show that they will not mix at any order of perturbation theory \cite{imbrie2016many,imbrie2017local}. By mixing we mean that the resonant pairs can exchange energy and become strongly entangled in the eigenstates. 
 The fact that resonances are rare and isolated at $L_W\geq 3.5$ is intimately related to the low, area-law entanglement scaling of eigenstates, and the existence of a complete set of LIOMs~\cite{Serbyn13-2,Serbyn13-1,Vosk13,Huse13,ScardicchioLIOM}. As the disorder strength is decreased, eventually the resonant pairs of spins become mixed, forming a connected network; then, LIOMs are destroyed, becoming nonlocal, and the system exits the MBL phase.

What is the proper {\emph{quantitative}} measure of disorder strength in an $SU(2)$ symmetric spin chain,  Eqs. (\ref{Eq:Ham}), (\ref{Eq:PJ})? The estimate (\ref{Eq:disorder})  for the {\it typical} ratio of the neighboring couplings in a Heisenberg chain suggests that the disorder experienced by the system becomes exponentially large in $1/\alpha$. Therefore, naively one could expect that, similar to the case of the random-field XXZ chain, where $L_W\propto W$, a lengthscale $L\propto e^{1/\alpha}$ (the inverse of the typical ratio of neighboring couplings) would determine the density of rare resonances. However, as we show in Sec. \ref{Sec:Resonances}, in fact, another measure of disorder is important. Specifically, one can introduce a lengthscale $L_1(\alpha)$ with the meaning similar to that of the length $L_W$ in the XXZ model (\ref{Eq:HXXZ}): $L_1(\alpha)$ defines a typical distance between local resonances in the system.  This lengthscale diverges when $\alpha$ goes to zero but, in contrast to the typical ratio of couplings, Eq. (\ref{Eq:disorder}),  only in a power-law fashion. Our numerical findings below are consistent with  $L_1(\alpha)\propto \alpha^{-0.4}$. 

 If the usual MBL scenario applied here, some $L_c=O(1)$ would exist such that, if $L_1(\alpha)\geq L_c$ the resonances  would not proliferate and the novel non-ergodic phase would be stable. 
Instead, the entanglement pattern of eigenstates, and the nonlocal nature of some integrals of motion induced by $SU(2)$  lead to the eventual proliferation of resonances at {\it any}  disorder strength, and so for {\it any} value of $L_1(\alpha)$, provided the system is sufficiently large.  Thus, another scale marking the crossover from the localized to the ergodic phase emerges. We denote this lengthscale, where ergodicity is restored, by  $L_{\rm erg}(\alpha)$.  In the subsequent Sections we provide strong evidence for the delocalization scenario described above. In systems with relatively weak disorder, length $L_{\rm erg}(\alpha)$ manifests itself e.g. in the level statistics and ETH violation  for matrix elements of  local observables that we study via exact diagonalization (see Sec. \ref{Sec:ShortScales} for details). At stronger disorder, no tendency towards ergodicity restoration can be observed in ED studies due to size limitations. However, a detailed analysis of the resonant processes  (see Sec. \ref{Sec:Resonances}) allows us to estimate   $L_{\rm erg}(\alpha)$ in this case as well.

\subsection{SDRG and excited eigenstates of the Heisenberg chain}
\label{Sec:RGandTreeStates_2}

In this Subsection, we qualitatively describe the SDRG approach to the disordered Heisenberg chains and discuss the properties of tree tensor-network states that it yields. A detailed description of the method is provided in the Appendix.   We emphasize that such states differ from the conventional MBL ones in two crucial aspects: first, they have a parametrically larger entanglement entropy, and second, one cannot define a complete set of LIOMs for them. 


A very large typical ratio of two neighboring couplings found for small $\alpha$, Eq.~(\ref{Eq:disorder}), suggests that the properties of the system can be described using the SDRG framework. The idea of SDRG is to identify a local ``grain'' in the system that is strongly coupled inside, but, due to strong disorder,  only weakly coupled to the rest of the system. The state of the grain is then approximated by one of the eigenstates of  its Hamiltonian, with the rest of the system decoupled. If one is looking for the ground state, the eigenstate of the grain is chosen to be its ground state.
Alternatively, if one aims to construct a random highly excited eigenstate that is effectively at an infinite temperature, as we are in this paper,  some eigenstate of the grain is randomly chosen. Then, the effective Hamiltonian of the system in which the grain is in the chosen eigenstate (or, more generally, a multiplet of states if symmetries dictate degeneracies in the spectrum of the grain's Hamiltonian)  is  calculated by perturbation theory in the grain-system coupling.  

\begin{figure}[t]
\includegraphics[width=250pt]{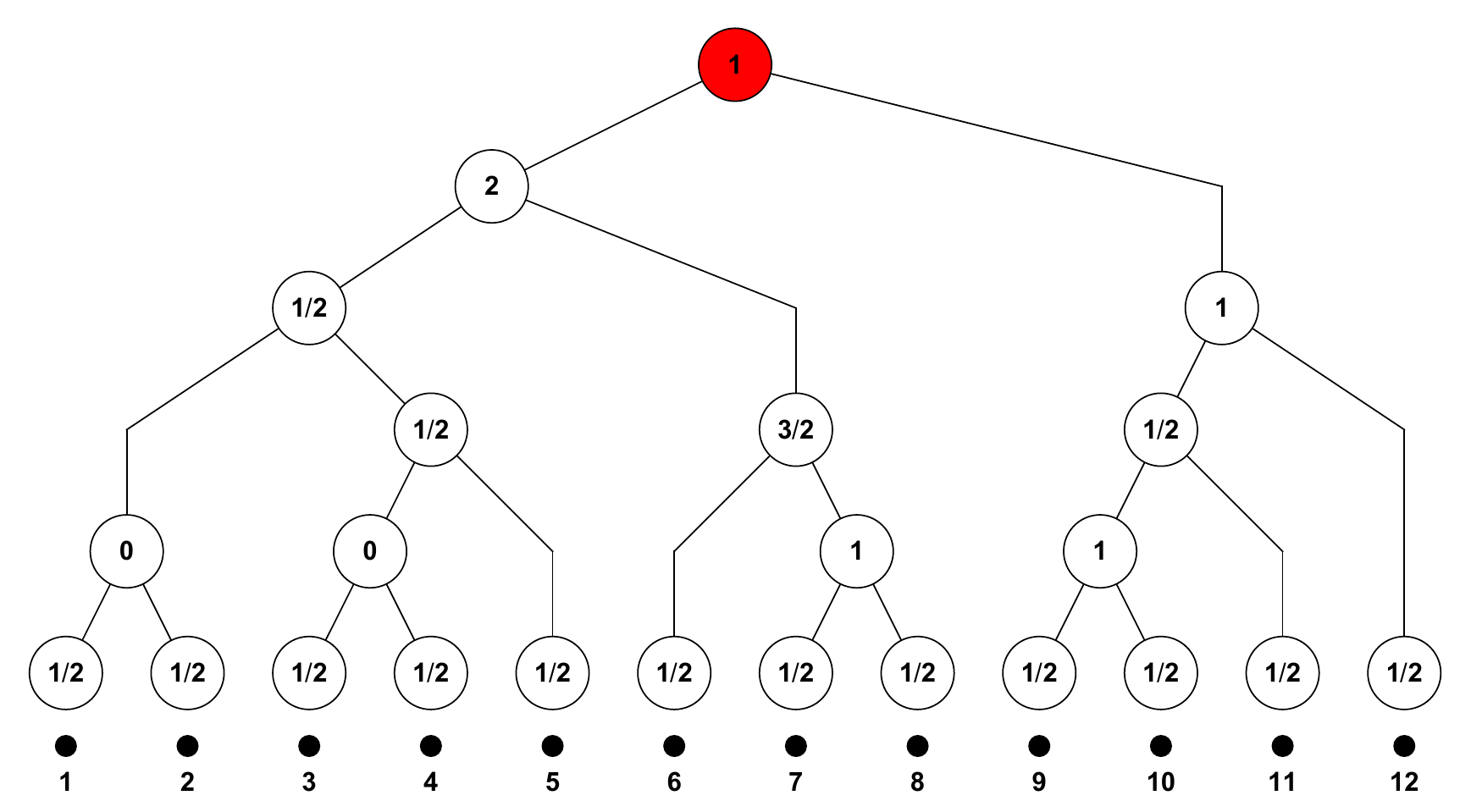}
 \caption{A multiplet of eigenstates predicted by the SDRG for a system of 12 spins 1/2. The leaves of the tree represent elementary spins in the system. The  tree describes the way the elementary spins are fused into larger block spins in the course of the SDRG. The numbers in the nodes indicate the resulting spins of the blocks.  The value in the top node (marked red) is the total spin $S_0$ of the system ($S_0=1$ in the present example). $S_0$ is an exact integral of motion. $(2S_0+1)$ different states in the multiplet can be distinguished by additionally  specifying the projection of the spin in the top node to the $z$-axis. 
 }
\label{Fig:tree}
\end{figure}

One can continue this procedure, assuming that the disorder in the effective Hamiltonian remains strong. This is indeed the case for e.g., ground states of random antiferromagnetic (AFM) Heisenberg chains~\cite{Fisher1992}. Then, a repeated application of the SDRG rules results in an approximate wave function of the whole system, obtained by ``patching" together the wave functions of the grains. 

A detailed discussion of the SDRG rules for excited states of the Heisenberg chain can be found in Ref.~\cite{Agarwal2015}, and we provide it in Appendix \ref{App:SDRG}. Qualitatively, for this system a grain is  a pair of neighboring spins coupled by a strong bond; its eigenstates (which come in $SU(2)$ multiplets) are labeled  by the total spin of the grain. The SDRG procedure replaces such spin pairs by effective (typically larger) spins, i.e. it assigns some total spin to larger and larger blocks of contiguous spins in the system. The resulting approximation for an eigenstate (more precisely, for a degenerate symmetry-enforced multiplet~\footnote{As most properties of the states comprising a multiplet are actually independent of the particular state and depend on the multiplet alone, throughout the manuscript we often refer to the SDRG trees as specifying a single quantum state. One may assume, for example, that in each of the multiplets we focus on a state with the $z$-projection of the total spin $0$ (1/2) if the length of the system is even (odd).}) is a kind of a tree tensor network, illustrated in Fig.~\ref{Fig:tree}. The nodes of the tree represent the block spins identified in the SDRG process.  The structure of the tree  reflects the order in which the elementary spins of the system should be added up to give an (approximate) eigenstate.

The fusion of spins in the course of the SDRG must be supplemented by a perturbative account of the interaction of merging spins with the rest of the system. 
In the present setting of an infinite-temperature SDRG, where spins typically fuse into non-singlet states, a first-order perturbation theory (that simply amounts to the projection of the fusing spins onto the direction of the total spin) suffices in most cases. The resulting renormalization of couplings is weaker than the one that occurs in the low-temperature SDRG for AFM spin chains, where the spins always fuse into singlets, and therefore a second-order perturbative treatment is required to find new renormalized couplings (see Appendix \ref{App:SDRG}). Still, the distribution of couplings developed in the course of SDRG turns out to be broad (see Ref.~\cite{Agarwal2015} and below).

Within the  SDRG approximation, the values of the block spins (the numbers associated  to the nodes of the tree in Fig. \ref{Fig:tree}) label the eigenstates of the Hamiltonian and bear similarity to the LIOMs of the conventional MBL phase. An eigenstate of the Hamiltonian is also an eigenstate of a sequence of these operators, just as an eigenstate of an MBL Hamiltonian is simultaneously an eigenstate of each LIOM. However, there are two major differences between these quantum numbers and LIOMs.

First, in the MBL phase the eigenstates of $H$ are at the same time eigenstates of \emph{a fixed set} of LIOMs. Total spins of the blocks in our problem would form conserved operators if different eigenstates were represented by \emph{geometrically identical} trees, which differ only in the values of the block spins. In reality, 
the order in which the spins are merged in the course of SDRG depends not only on the particular disorder realization, but also on the eigenstate of the grain, which is randomly picked at any given step of the SDRG (see Appendix \ref{App:SDRG} for details). Thus, the values of the block spins, in general, cannot be promoted from labels of a particular eigenstate to operators acting in the full Hilbert space. The structure of larger blocks depends on the history of choosing total spins at the earlier steps of SDRG. 

Second, LIOMs in an MBL system are quasi-local, exponentially localized in space operators~\cite{Serbyn13-2,Serbyn13-1,Huse13}. In contrast, the block spins of the strongly disordered Heisenberg chain have a hierarchical structure. While some of them (living near the bottom of the tree) can be expressed in terms of an $O(1)$ number of the original spin operators $\mathbf{S}_i$,  the other ones, found at the higher levels of the tree, are highly nonlocal in terms of the original spins. Thus, $SU(2)$ symmetry forces some integrals of motion to become nonlocal. Therefore, SDRG (in the regime of its validity) describes a non-ergodic phase of a new kind, with a partial, rather than complete set of LIOMs. Our goal is to investigate the stability of this putative phase.

The novel non-ergodic character of tree eigenstates manifests itself in the scaling of entanglement entropy. For simplicity, we will consider the entanglement entropy of an eigenstate with respect to the a cut in the  middle of the chain,
\begin{equation}
S_{\rm ent}(L/2)=-\mathrm{Tr}(\rho_{L/2}\log_2\rho_{L/2}),
\label{eq:entS}
\end{equation}
where $\rho_{L/2}$ is the reduced density matrix of half-chain in the chosen eigenstate and the trace is taken over the degrees of freedom in the other half of the system.   
A bound for the entanglement entropy depends on the tree structure describing a given state, in particular, on the tree depth $d$  (the number of levels between the very top node of the tree and the original physical spins). We find, via numerical simulations, that typical states produced by the SDRG procedure have a logarithmic depth, $d\propto \ln L$.  It is then  possible to show (see Appendix \ref{App:Entanglement}) that the entanglement entropy of a single  typical~\footnote{The upper bound on the entanglement  entropy in Eq. (\ref{Eq:SUp}) holds in fact for all the states described by trees of logarithmic depth, see Appendix \ref{App:Entanglement}.} tree satisfies
 \begin{equation}
c_1 \log_2 L\lesssim S_{\rm ent}(L/2)  < c_2 \log_2^2 L,
 \label{Eq:SUp}
 \end{equation}
 where $c_1$ and $c_2$ are numerical constants of order unity that depend on the statistical properties of the tree.
Thus, the entanglement of the tree states scales faster than the area-law found in MBL, but significantly slower compared to the thermal entanglement for an infinite-temperature state,  $S_{\rm th}(L/2) \approx \frac{L}2$ (measured in bits). 

The upper bound on the entanglement entropy in Eq. (\ref{Eq:SUp}) can also be generalized (see Appendix  \ref{App:Entanglement}) to the case when the state in question is not a single tree state but rather a linear combination of $n_{\rm T}$ tree states:
\begin{equation}
S_{\rm ent}(L/2)  < c_2 \log_2^2 L +\log_2 n_{\rm T}.
\label{Eq:SUpLin}
\end{equation}
Although this bound might seem weak, it has an important implication, which will be used below: if the system's eigenstates become ergodic, they must be represented by an exponentially large number of tree states. 

\subsection{Validity of SDRG and the (in)stability of tree states}
\label{Sec:RGandTreeStates_3}

The SDRG is a heuristic procedure relying on strong disorder. The tree states generated by SDRG are not exact eigenstates of the Heisenberg spin chain, but how accurate are they? Historically, at each step of SDRG one checks that the disorder in the effective Hamiltonian remains strong, such that strong couplings can be found; one can then check for the absence of resonances involving a small number of spins, to make sure that the neglected processes do not destroy the tree structure. While for the analysis of ground states this is often sufficient, it is unclear whether such tests can guarantee the accuracy of SDRG for the excited states. 

Below we will check the validity of SDRG for excited states using several approaches. First, we will compare SDRG tree states to the exact eigenstates for system sizes up to $L=26$, obtained numerically. We will use a number of measures, such as level statistics, and the eigenstate thermalization hypothesis (ETH) and its breakdown. 
Second, to describe large system sizes, we will develop an approach to account for many-body processes that are usually neglected in SDRG, and to test their relevance. We introduce this approach qualitatively now, and we will apply it in what follows. Suppose that SDRG yielded some tree state $|\Psi^0_{\rm RG}\rangle$, specified by the tree geometry and the choice of total spins in each node. Instead of considering the effective Hamiltonian at every step, we can write the original Hamiltonian exactly in the basis of tree states with the geometry identical to that of $|\Psi^0_{\rm RG}\rangle$. The first key observation is that the selection rules imposed by symmetry facilitate the analysis of relevant processes; more specifically, the block spins along a path connecting a pair of contiguous physical spins to the top of a tree can change by $\Delta S=0,\pm 1$~\footnote{There are some additional restrictions. Specifically, the new set of block spins in the tree should still be consistent with the rules of angular momentum addition. In particular, all the new block spins should be non-negative.  Moreover, in certain cases the transitions with $\Delta S=0$ are forbidden  in full analogy with quantum optics, where the  $\Delta l=0$ transitions are forbidden for initial (or final) state with an angular momentum $l=0$, see for details Ref.~\cite{Protopopov2017} and Appendix \ref{App:SDRG}}.  The second observation is that, given that the typical spins of larger blocks grow (as a square root of the block size) the tree states connected to $|\Psi^0_{\rm RG}\rangle$ by the Hamiltonian are expected to have the same geometrical structure. This is because for large spins, strong bonds remain strong when a value of some spins is changed by $\Delta S\ll S$. 

We search for {\it resonances} between different tree states and characterize their properties. Solving the full eigenvalue problem for large $L$ is hopelessly complicated; thus, we focus on low-order resonances. Effectively, we check whether the Hamiltonian hybridizes a given tree state with its neighbor, say $|\Psi^1_{\rm RG}\rangle$ (a neighbor is a state such that $\langle \Psi^1_{\rm RG}| H|\Psi^0_{\rm RG}\rangle\neq 0$). As long as the probability of finding resonances is sufficiently low, we expect that true eigenstates are localized in the tree basis. This corresponds to a non-ergodic phase, or regime (if it occurs only for sufficiently small system sizes). Alternatively, if there are many resonances which proliferate, it is natural to expect that the SDRG breaks down and the system become ergodic. 

It is instructive to draw parallels with the conventional MBL phase of the strongly disordered  XXZ spin chain in a random magnetic field. The caricature of MBL eigenstates is just product states with a well defined $S_i^z$ projection for each spin. While corrections to this picture certainly exist (e.g. LIOMs are not strictly equal to $S_i^z$ operators) we know that  MBL is stable, if the disorder is sufficiently strong. Our aim is to understand whether for $SU(2)$-symmetric chains tree states, with their built-in correlations and unusual entanglement  properties, can be stable, representing a dynamical phase distinct from both MBL and ergodic phase.

Below we will use the above aproach to reveal a broad non-ergodic regime where tree states are stable. We will also provide evidence that trees eventually become unstable above certain system size (dependent on $\alpha$) for all values of $\alpha$ that we study. We therefore propose the picture that, while for finite systems the dynamics is non-ergodic at strong disorder, in the thermodynamic limit, ETH should be recovered (see Fig.~\ref{Fig:schematic}).

\section{Exact diagonalization studies}
\label{Sec:ShortScales}

In this Section, we present our numerical results from exact diagonalization.

\subsection{Probing the stability of tree states}

\label{Sec:ShortScales_2}

To analyse the accuracy of the SDRG procedure, we first study the participation ratios of exact eigenstates of the system (\ref{Eq:Ham}) in the basis of the tree states generated by SDRG. More precisely, for a given disorder realization $\{J_i\}$, we first run the  SDRG to generate some tree state $\left|\Psi_{\rm RG}^0\right\rangle$ with a total spin $S_0$. A complete basis of states in the sector with a given total spin $S_0$ (and some fixed $z$-projection of the total spin)  can be built out of $\left|\Psi_{\rm RG}^0\right\rangle$ by fixing the geometry of the underlying tree, but allowing the block spins in the tree (apart from the top one, $S_0$) to take all possible values consistent with the angular momentum addition rules. 
We denote the basis obtained in this manner by
\begin{equation}
\{|\Psi^a_{\rm RG}\rangle\}_{a=0,...,\mathcal{D}_{S_0, L}-1},
\label{eq:SDRGbasis}
\end{equation}
where $L$ is the length of the chain and  $\mathcal{D}_{S_0, L}$ is the Hilbert space  dimension of the sector with a total spin $S_0$, and a fixed projection $S_z=0$,
\begin{equation}
 \mathcal{D}_{S_0, L}=
 C_{L}^{L/2+S_0}-C_{L}^{L/2+S_0+1},
 \qquad  C_n^m\equiv\frac{n!}{m!(n-m)!}.
 \label{Eq:D}
 \end{equation}

The state with an index $a=0$ is the  original SDRG state, $\left|\Psi_{\rm RG}^0\right\rangle$. In general,  due to the correlations between the geometric structure of the tree and the values of the block spins discussed in Sec. \ref{Sec:RGandTreeStates_2},  many of the states in the basis (with indices $a>0$)  would not be approximate eigenstates constructed in RSRG. We expect, however, that at strong disorder the geometry of the tree that corresponds to the state $\left|\Psi_{\rm RG}^0\right\rangle$ is also appropriate for a  number of other SDRG states that do not differ too much from $\left|\Psi_{\rm RG}^0\right\rangle$ in the values of block spins. In that case, a significant fraction of $\left|\Psi_{\rm RG}^{a>0}\right\rangle$  {\emph are} in fact ``SDRG eigenstates''.

We then perform an exact diagonalization of the Hamiltonian in the basis (\ref{eq:SDRGbasis})~\footnote{Explicit expressions for the matrix elements of the operators ${\bf S}_{i}\cdot {\bf S}_{j}$ were derived in Ref.~\cite{Protopopov2017}}.  Among all the eigenstates of the Hamiltonian we focus on a single one, denoted by  $|E\rangle$, that has a maximum overlap with a given $|\Psi^0_{\rm RG}\rangle$. 
The quality of $|\Psi^0_{\rm RG}\rangle$ as an approximation to  $|E\rangle$ can  be quantified
by the  inverse participation ratio (IPR) of the state $|E\rangle$ in the SDRG basis (\ref{eq:SDRGbasis}):
\begin{equation}
I_{E}=\sum_{a=0}^{\mathcal{D}_{S_0, L}-1}\left|\langle E| \Psi_{\rm RG}^a\rangle\right|^4,
\label{Eq:IPR}
\end{equation}
and its inverse $N_{E}\equiv 1/I_E$ which can be viewed as the number of tree states $\left |\Psi_{\rm RG}^a\right\rangle$ (of a given topology) that one needs to represent the eigenstate~$|E\rangle$. Thus, small values of $N_E\sim 1$ indicate that the SDRG is accurate, while very large $N_E\gg 1$ signals an instability of tree states.

\begin{figure}
\includegraphics[width=210pt]{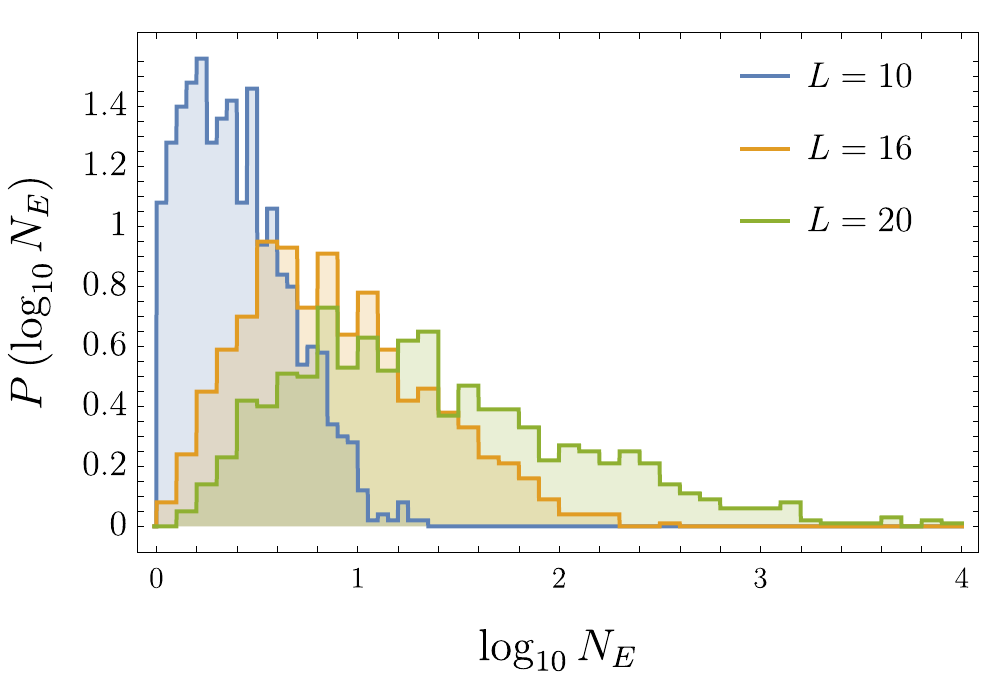}
\includegraphics[width=210pt]{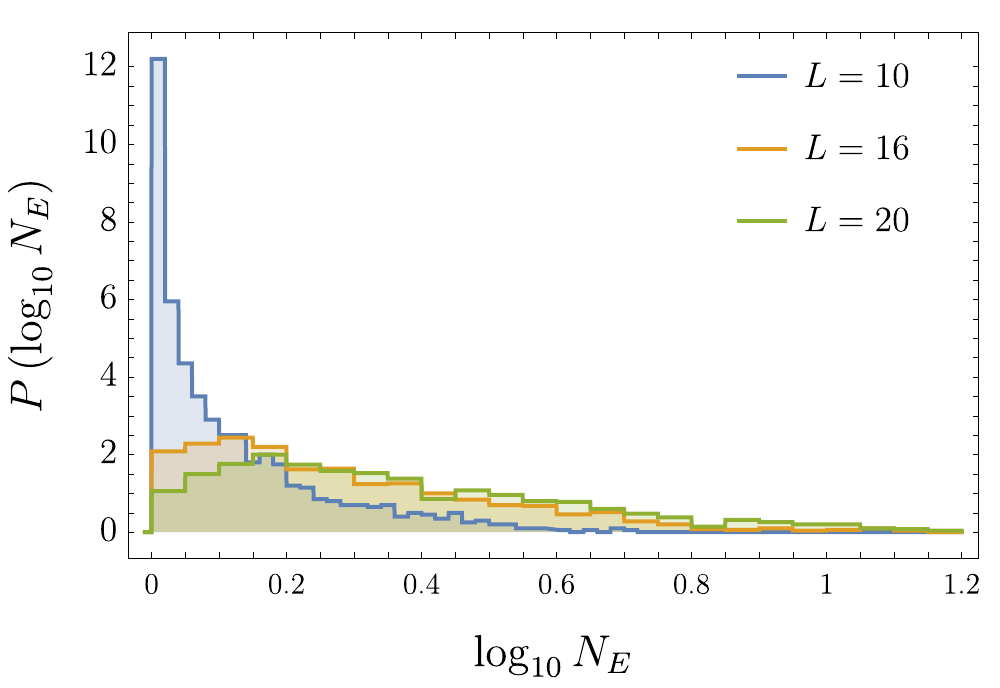}
\caption{Statistics  of $\log_{10} N_E$ for $\alpha=1$ (upper panel) and $\alpha=0.3$ (lower panel). 
Different curves correspond to different system sizes, $L=10$,  $L=16$ and $L=20$, see legend. 
}
\label{Fig:IPRDistribution}
\end{figure}

\begin{figure}
\centering
\includegraphics[width=230pt]{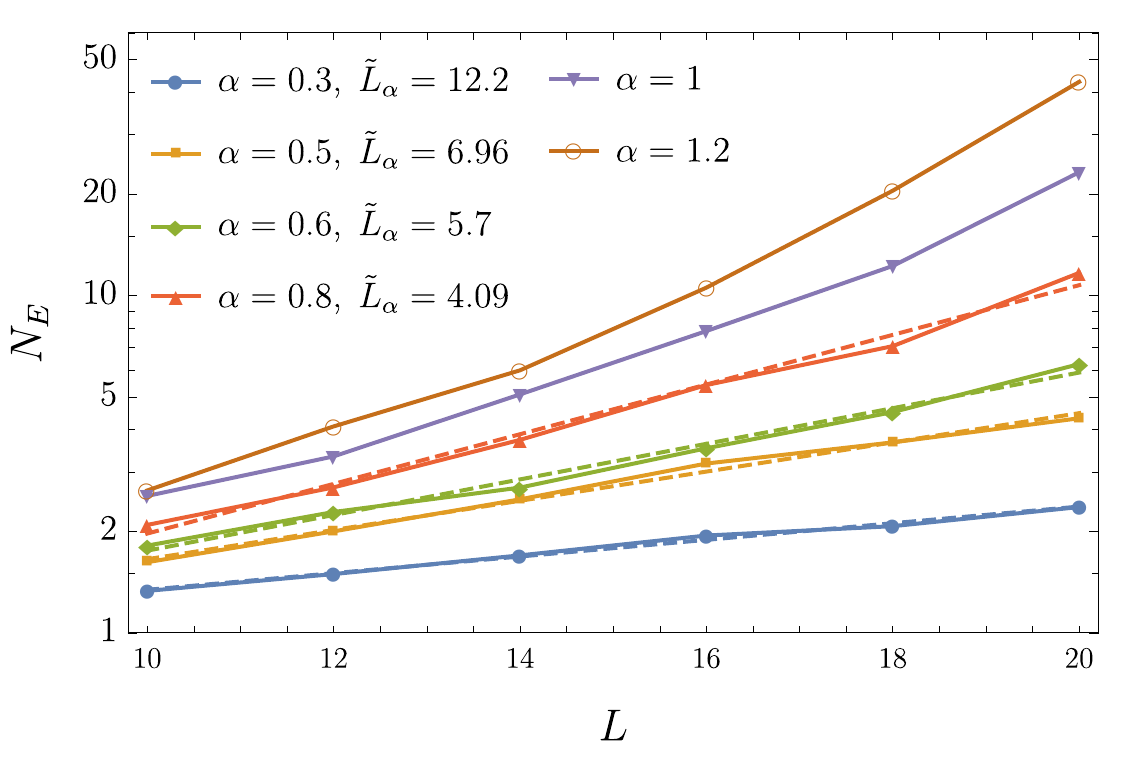}
\includegraphics[width=230pt]{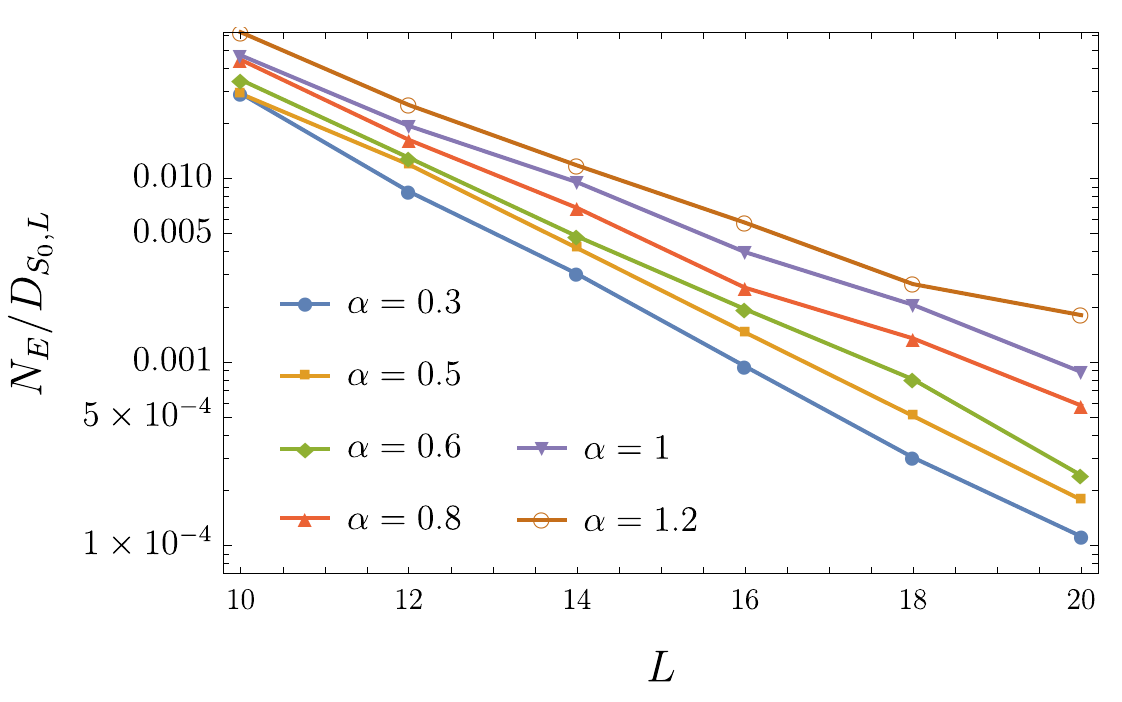}
\caption{ Typical number $N_E$ of tree states participating in the eigenstate $|E\rangle$ (top) and 
the typical value of the  fraction $N_E/D_{S_0, L}$ (bottom) versus the system length for different strengths of disorder (see legend). 
The dashed lines in the top panel represent the exponential fits $N_{E}\propto 2^{L/\tilde{L}_1(\alpha)}$.  }
\label{Fig:IPRMean}
\end{figure}

Computing the participation ratio $N_E$ for $10^3$ disorder realisations $\{J_{i}\}$ (and a single random SDRG state $\left|\Psi^0_{\rm RG}\right\rangle$ for each  $\{J_{i}\}$), we investigate the statistical properties of this quantity.      
We performed numerical simulations for the disorder parameter $\alpha$ ranging from $\alpha=1.2$ (weak disorder) to $\alpha=0.3$ (strong disorder). The results are summarized in Figs.~\ref{Fig:IPRDistribution} and \ref{Fig:IPRMean}. Figure~\ref{Fig:IPRDistribution}  shows several examples of the distributions  of $\log_{10} N_E$ for different system sizes and two different disorder strengths.  We observe that in short systems, $L=10$, $|\Psi^0_{\rm RG}\rangle$ is very close to an exact eigenstate even for weak disorder, $\alpha=1$, in the sense that   $N_E\sim 1$. Upon increasing  the system size $N_E$ grows, signalling that approximating the eigenstate $|E\rangle$ with a tree state  $|\Psi^0_{\rm RG}\rangle$ becomes less accurate.  

The evolution of the typical value of $N_E$ (defined as $e^{\langle \ln N_E\rangle}$) with the system size is illustrated in the top panel of Fig. \ref{Fig:IPRMean}. Interestingly, even in the weak disorder regime, $\alpha=1$, and for the largest system size $L=20$, the typical $N_E\sim 25$ remains small compared to the dimension of the Hilbert space $\mathcal{D}_{S_0, L}$.  The latter depends on the spin sector $S_0$, which is chosen at random in the present  analysis.  The SDRG procedure we use generates states with different $S_0$ in accordance with their probability in the infinite temperature ensemble, $P(S_0)\propto (2S_0+1){\cal D}_{S_0,L}$. 
For $L=20$ the most frequently encountered value of $S_0$ is $3$, corresponding to the Hilbert space dimension  $\mathcal{D}_{3, 20}=38760$. Moreover, for $90\%$ of the SDRG states $S_0\leq 5$ and ${\cal D}_{S_0, 20}\geq 10659$. The length dependence of the typical Hilbert space fraction occupied by the energy eigenstate $|E\rangle$ (in the tree basis), $e^{\langle \ln N_E/\mathcal{D}_{S_0, L} \rangle}$, is shown in the bottom panel of Fig.~\ref{Fig:IPRMean}.


It is instructive to compare the above findings to the behavior of IPR in the product state basis for the conventional MBL phase. Viewing MBL as a kind of Anderson localization in the Hilbert space, one might naively expect that in the MBL regime the eigenstates would exhibit system-size independent  IPR,  $N_E\gtrsim 1$.  It is known~\cite{Bauer13, Luca13, Alet14, ALET2018498}, however, that in reality MBL eigenstates are rather {\it fractal} when viewed in the product-state basis: the participation ratio $N_E$ scaling as $N_E\propto\mathcal{D}^{\gamma}\propto 2^{\gamma L}$ with an exponent that depends on disorder strength. The fractal behavior stems from perturbative corrections, and resonances discussed  at the end of Sec. \ref{Sec:RGandTreeStates_1} (or, equivalently, it is due to the fact that local integrals of motion have support over more than one lattice site). In the strong disorder limit, $\gamma \propto t/W=1/L_W\ll1$.  The MBL transition is thus marked {\it not} by the emergence of the growth of $N_E$ with the system size, but rather by a jump of the exponent $\gamma$ to its thermodynamic value, $\gamma=1$ (at infinite temperature).

The behavior shown in Fig. \ref{Fig:IPRMean} for the Heisenberg chain is qualitatively similar. At strong disorder, $\alpha \leq 0.8$,  the dependence $N_E(L)$ for the available system sizes  can be approximated by an exponential fit, $N_E\propto 2^{L/\tilde{L}_1(\alpha)}$ (see dashed lines in the top panel of Fig.~\ref{Fig:IPRMean}; the corresponding values of the fitting parameter $\tilde{L}_1(\alpha)$ are indicated in the legend). The length $\tilde{L}_1(\alpha)$ grows as the disorder strength is increased.  By analogy with the conventional MBL, we can expect that lengthscale $\tilde{L}_1(\alpha)$ characterizes the  density $1/\tilde{L}_1(\alpha)$  of the rare  local resonant degrees of freedom in the system, see also the discussion at the end of Sec. \ref{Sec:Resonances_2}.

At weaker disorder, $\alpha=1,\, 1.2,$ the naive exponential fit would produce a very small $\tilde{L}_1(\alpha)<2.5$. Moreover, the slope $d\ln N_{E}/dL$ of the corresponding lines shows a clear increase as the system size grows. Accordingly, the fraction $N_E/D_{L, S_0}$ (bottom panel of Fig.  \ref{Fig:IPRMean} ) displays a tendency towards saturation suggesting that ultimately, the scaling of $N_E$ in long systems becomes ergodic, $N_E\propto 2^L$.



In view of the results above, it may be tempting to conclude that the strongly disordered Heisenberg spin chains  do indeed display a non-ergodic, non-MBL phase with unusual tree-like eigenstates that are only slightly dressed by perturbative corrections and occasional resonances (similar to how in the conventional MBL phase the eigenstates are perturbatively dressed product states). At weaker disorder, one would then expect a transition into an ergodic phase. However, the crucial question concerns the ultimate fate of the putative non-ergodic behavior in the thermodynamic limit. In particular, does the observed fractal scaling, $N_E\propto 2^{L/\tilde{L}_1(\alpha)}$ persist, or does it eventually cross over to the ergodic scaling, as for the weakly disordered case?  In order to answer these questions, in the next Sections we will subject the hypothetical non-ergodic phase to several stringent tests.


\subsection{Level statistics}
\label{Sec:ShortScales_1}

\begin{figure}
\includegraphics[width=220pt]{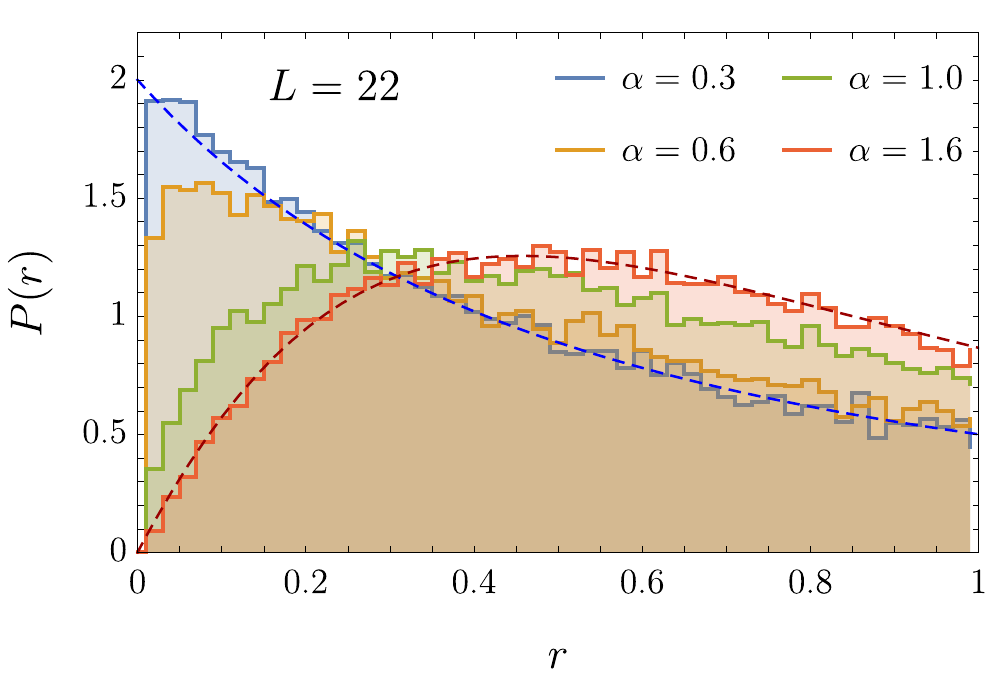}
\includegraphics[width=220pt]{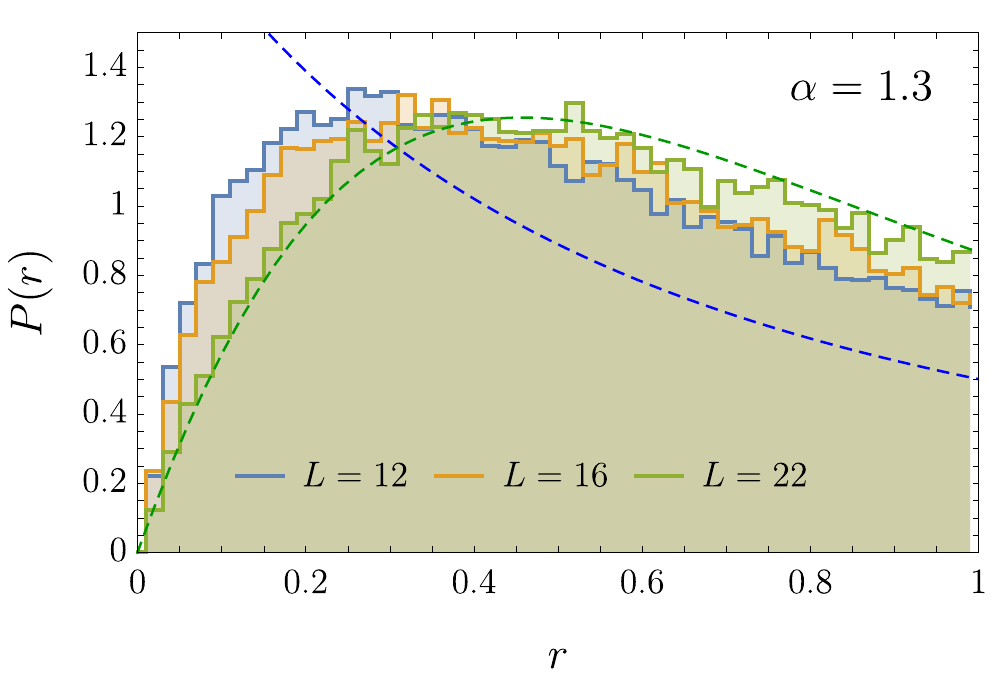}
\caption{Level statistics for the Heisenberg chain. Each curve in the figure was produced using at least 50 eigenstates from the middle of spectrum and 1000 realizations of disorder. The dashed lines are the Wigner-Dyson (WD) and Poisson distributions. (Top) For a fixed length $L=22$ and varying $\alpha$. (Bottom) For a fixed $\alpha=1.3$ and varying length $L$. The tendency towards the Wigner-Dyson statistics is evident both at growing $L$ and $\alpha$. However, for smaller values of $\alpha$, $P(r)$ remains close to the Poisson one up to largest available system sizes.}
\label{Fig:Spectra}
\end{figure}
\begin{figure}
\includegraphics[width=0.95\columnwidth]{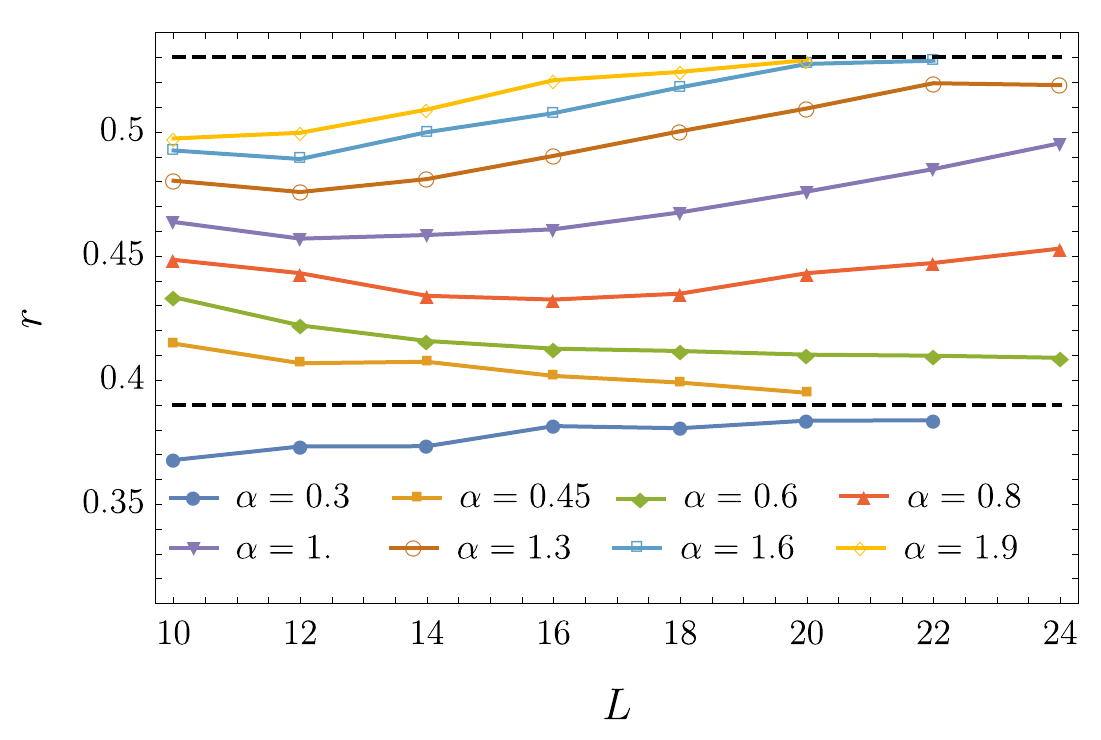}
\caption{The average $r$ parameter as a function of $L$ for different values of $\alpha$. From top to bottom (for any fixed $L$) $\alpha=1.9,1.6,1.3,1,0.8,0.6,0.45,0.3$. The dashed lines at $r=0.53$ and $r=0.39$ represent the WD and Poisson values, respectively. Error bars are within the symbol. 
}
\label{Fig:RMean}
\end{figure}
\begin{figure}
\includegraphics[width=0.95\columnwidth]{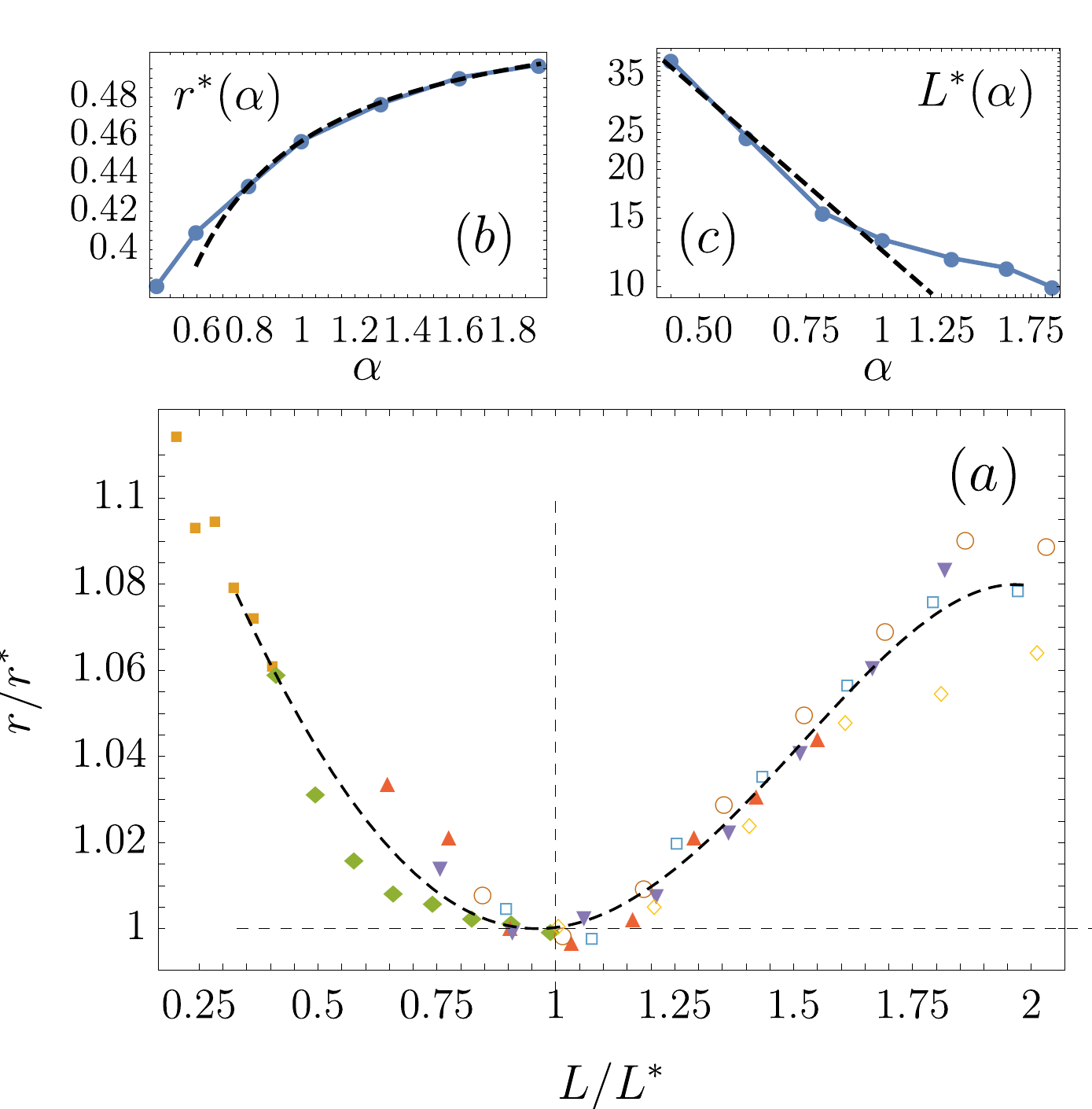}
\caption{(a) The average $r$ parameter as a function of $L/L^*$, divided by $r^*$, falls onto a universal curve in the vicinity of its minimum; (b) The value of the minimum $r^*(\alpha)$. The dashed line is a fit of the form $r^*(\alpha)=r_\infty +c_1/\alpha+c_2/\alpha^2,$ which returns $r_\infty=0.53\pm 0.01$ compatible with the GOE value; (c) The position of the minimum $L^*(\alpha)$ for $\alpha=1.9,1.6,1.3,1,0.8,0.6,0.45$. The dashed line is  a fit of the form $L^*(\alpha)=c\ \alpha^{-\nu}$ with $\nu=1.4\pm 0.13$ for the first 4 points.
}
\label{Fig:LstarRstar}
\end{figure}

Our main goal in this Subsection is to further characterize non-ergodic behavior found above, and its dependence on the system size. We will employ the standard diagnostic of ergodicity and its breakdown: the level statistics in the center of the many-body band. An extensive use of the constraints imposed by $SU(2)$ symmetry allows us  to perform exact diagonalization on spin chains of up $L=26$ spins. Larger system sizes that we can achieve here compared to Subsection \ref{Sec:ShortScales_2} are due to the use of massively parallel algorithms together with the possibility to focus on a small number of eigenstates near the band center (recall that the identification of the eigenstate $|E\rangle$ studied in Sec.~\ref{Sec:ShortScales_2} required the knowledge of the full set of eigenstates).  
In most of our studies, we concentrated on the $S_0=0$ sector, and data for $S_0=1,2$ did not show any qualitative differences. For each $L,\alpha$ and each disorder realization $\{J_i\}_{i=1,...,L}$, up to 50 eigenstates around the middle of the spectrum (fewer for $L=10,12$ and $L=26$) were obtained, and a total of at least $1000$ disorder realizations (except for $L=26$) where considered. 

We characterize the level statistics by the $r$-parameter, defined as follows~\cite{OganesyanHuse}:
\begin{equation}
r\equiv\frac{\min(\Delta_n, \Delta_{n+1})}{\max(\Delta_n, \Delta_{n+1})},
\end{equation}
with $\Delta_n$ and $\Delta_{n+1}$ being two consecutive level spacings.

The distribution of the parameter $r$ and its dependence on the system size and disorder strength are shown in  Figs.~\ref{Fig:Spectra} and~\ref{Fig:RMean}. The distributions of $r$ change qualitatively as $\alpha$ is decreased at a fixed $L$: for largest $\alpha=1.6$ (very weak disorder), $r$ is described by the standard Wigner--Dyson distribution, while for small $\alpha=0.3$ (strongest disorder considered) one observes the Poisson distribution, with virtually no level repulsion. This supports the existence of a non-ergodic regime at accessible system sizes. For $\alpha\in [0.6;1]$, the level statistics is intermediate between the Wigner--Dyson and Poisson distributions. We also illustrate the dependence of the distribution $P(r)$ on the system size for weak disorder $\alpha=1.3$. It is evident that the distribution flows towards Wigner--Dyson, albeit relatively slowly. 

Further, we study the flow of the average value, $\langle r\rangle$, with the system size, in an attempt to extract some relevant lengthscales.  $\langle r\rangle$ as a function of $L$ for different values of $\alpha$ is illustrated in Fig.~\ref{Fig:RMean}. For the weak disorder, $\alpha\geq 0.8$, 
 the dependence of $\langle r\rangle$ on $L$ is non-monotonic. 
Our data show a tendency towards the Poisson statistics for small system sizes, $L<L^*(\alpha)$, but for $L>L^*(\alpha)$ the value of $\langle r \rangle$ starts growing, moving towards the Wigner--Dyson (WD) value. Upon decreasing $\alpha$ to the value of $0.8$, the lengthscale $L^*(\alpha)$  increases, while its value $r^*(\alpha)\equiv r\left[L^*(\alpha)\right]$ decreases. The ultimate flow of $\langle r\rangle$ towards the WD value  is consistent with the expectation that at weak disorder the system becomes ergodic for modest system sizes. One can  estimate the scale where system becomes ergodic, $L_{\rm erg}$,  by extrapolating the $\langle r(L)\rangle$ dependence till the crossing with the WD line.  The lengthscale extracted in this way is larger than the maximum system sizes accessible numerically for $\alpha<1$.  The extrapolation procedure suffers from a large uncertainty. Therefore, we chose instead to characterize the delocalization crossover by the length $L^*(\alpha)$, and we expect that $L_{\rm erg}(\alpha)\propto L^*(\alpha)$. 

 The data at stronger disorder, $\alpha\in [0.3;0.6]$ shows \emph{prima facie} a qualitatively different behavior. For the strongest disorder, $\alpha=0.3$, the parameter $\langle r \rangle$ slowly increases for small $L$, in a stark contrast with the behavior found for $\alpha\geq 0.8$. Interestingly, at small $L$ this parameter is below the Poisson value of $\langle r\rangle_{\rm P}\approx 0.39$. We attribute this to strong disorder leading to the appearance of very small couplings in a typical disorder realization (smaller than the level spacing at small $L$). The chain is then effectively broken into smaller, almost non-interacting, spin chains.  This leads to level clustering and the $r$ parameter becomes sub-poissonian. However, since the level spacing decreases exponentially with the system size, while the weakest coupling only decreases as a power-law (see Eq.~(\ref{Jmin})), the level clustering is eventually washed out and  for $L>18$ the parameter $\langle r \rangle$ rapidly approaches the standard Poisson value. For disorder strength $\alpha=0.45,0.6$, $\langle r \rangle$ is initially slightly above the Poisson value, but it {\it decreases} as the system size is increased; no flow towards WD is seen. For the system sizes analyzed, it is evident that ergodicity has not developed and a single SDRG tree state provides a good approximation to the eigenstates, as we also demonstrated in the previous Subsection.  

The exact diagonalization results for strong disorder values, $\alpha\in [0.3;0.6]$, may be consistent with two scenarios. One scenario is that (much like in the usual MBL) the system experiences a phase transition at some critical disorder strength. Another scenario is that, even at strong disorder, the system would eventually flow to ergodicity, similar to what we found for weaker disorder values. Assuming that this second scenario is realized, in large enough systems  the curves for $\alpha=0.45,0.6$ would first develop a minimum and then flow to the WD value at yet larger system sizes.  The corresponding scale $L^*$ can be heuristically extracted by extrapolating~\footnote{The extrapolation is performed and the value of $L^*$ is extracted by fitting the available data for $L\geq 14$ by a quadratic dependence, $r(L)=r^*+a (L-L^*)^2$ , with fitting parameters $r^*$, $L^*$ and $a$.  }
 the ED data shown in Fig.~\ref{Fig:RMean}. The dependence of the length $L^*$ on disorder, as extracted by the analysis outlined above, is illusrated in Fig.~\ref{Fig:LstarRstar}(c). It is consistent with a power-law scaling, $L^*(\alpha)\propto \alpha^{-1.4}$. We note that the curves $\langle r  (L) \rangle$ for $\alpha\geq 0.6$ (including extrapolated data for $\alpha=0.45,0.6$) can be collapsed (in the vicinity of $L=L^*$) into a single one by simultaneous rescaling $r\rightarrow r/r^*$ and $L\rightarrow L/L^*$, see Fig. \ref{Fig:LstarRstar}. 

To sum up, the length $L^*$ beyond which the spectral parameter starts flowing towards the WD value (but the system of size $L^*$ is still non-ergodic, because $r^*$ is closer to the Poisson value), grows rapidly with the increase of disorder. Although the trend is clear,  we are extrapolating significantly away from the accessible system sizes, $L\leq 26$. Thus, the law governing $L^*(\alpha)$ which we propose should be taken with a grain of salt. In the next Subsection, we proceed to test the eigenstate thermalization hypothesis. 

\subsection{Eigenstate thermalization hypothesis and its breakdown}
\label{Sec:ShortScales_ETH}

Next, we characterize the eigenstates of random Heisenberg chains by testing the the Eigenstate Thermalization Hypothesis (ETH) and its breakdown. ETH provides a microscopic picture of thermalization in ergodic quantum systems~\cite{SrednickiETH,Srednicki96, DeutschETH}. Specifically, it states that individual ergodic eigenstates appear to be thermal, from the point of view of simple physical observables (e.g.\ few-body operators).  ETH formalizes and extends the intuition that the eigenstates of an ergodic system should be ``as random as possible'', up to a small set of global constraints (in our case, energy and total spin).

For our purposes, ETH can be formulated in terms of the expectation values of local observables. Let $ \hat{O} $ be an operator representing some physical observable. Then, for every pair of eigenstates $|a\rangle, |b\rangle$ of an ergodic system, ETH yields an ansatz for matrix elements of $\hat{O}$~\cite{SrednickiETH,Srednicki96}: 
\begin{equation}\label{ETH-expvalue}
  \langle a | \hat{O} | b \rangle = \bar{O}(E)\delta_{ab} + e^{-S(E)/2} f(E, \Delta E) R_{ab},
\end{equation}
where $ E $ and $ \Delta E $ are, respectively, the average and the difference between the energies of the two eigenstates, $ E=\frac{E_a+E_b}2, \Delta E=E_b-E_a$, and $ S(E) $ is the microcanonical entropy. Function $ f(E,\Delta E)$ is a smooth function of its arguments, which reflects dynamical properties of observable $\hat{O}$ and is system-specific. Finally, $ R_{ab} $ is a normally distributed random variable with unit variance. Notably, in this formula the diagonal part $ \bar{O} $ is assumed to be a smooth function of $ E $ alone, and is equal to the microcanonical average of $\hat{O}$. This reflects the fact that observables in eigenstates are equal to their microcanonical ensemble values.

According to Eq.~(\ref{ETH-expvalue}), in a thermalizing system the distribution of values $  \langle a | \hat{O} | a \rangle $ for eigenstates $ |a\rangle$ that are sufficiently close in energy should display a reasonably smooth dependence on $ E $, with only small, normal fluctuations about the average, suppressed exponentially in the system size by the factor $e^{-S(E)/2}$ so as to reproduce the microcanonical ensemble in the infinite-size limit.

We have focused our attention on the following two local observables:
\begin{align*}
   \hat{O}_\mathrm{max} \equiv \mathbf{S}_{i^\star}\cdot\mathbf{S}_{i^\star+1},\\
   \hat{O}_\mathrm{rand} \equiv \mathbf{S}_{j^\star}\cdot\mathbf{S}_{j^\star+1},
\end{align*}
where $ (i^\star,i^\star+1) $ is the pair of spins coupled the most strongly ($ |J_{i^\star}| = \max|J_i| $) and $ (j^\star,j^\star+1) $ is its antipodal pair ($ j^\star = i^\star + L/2 $ modulo $ L $). Since couplings are independent, the latter pair is coupled by an interaction $ J_{j^\star} $ of typical (or ``random'') strength, hence the name $ \hat{O}_\mathrm{rand} $.

\begin{figure}
\includegraphics[width=1.0\columnwidth]{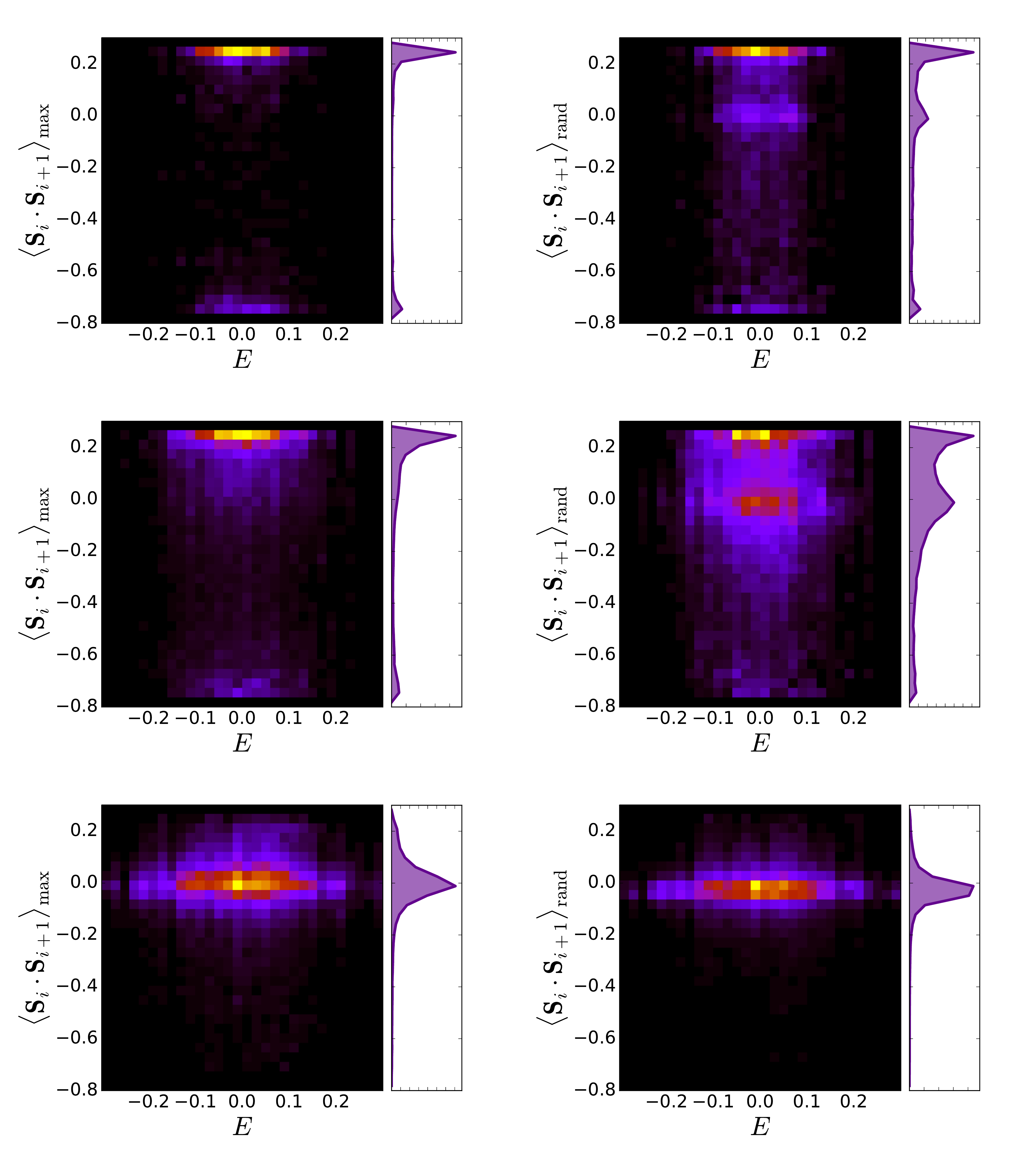}
\caption{Heat maps for the distributions of $ \langle a|\hat{O}_\mathrm{max}|a \rangle $ (left) and $ \langle a|\hat{O}_\mathrm{rand}|a \rangle $  (right) over several ($ \gtrsim 25000 $) eigenstates. Here $ S_0 = 0 $, $ L = 22 $ and $ \alpha \in \{0.3,0.6,1.3\} $ (top to bottom).The concentration of the \( y \)-marginals around 0 denotes increasingly ergodic behavior (see the comments in the text).}
\label{Fig:ETH-heatmaps}
\end{figure}

Let us discuss our expectations for the averages of these operators over eigenstates, depending on whether SDRG is accurate. 
First, suppose that $|a\rangle$ is exactly an SDRG tree state. Then the spins $(i^\star, i^\star+1)$ are going to be paired in either a $ S = 0 $ or a $ S = 1 $ state, and the value of $\langle a|\hat{O}_\mathrm{max}|a \rangle$ is going to be either $-3/4$ or $1/4$, respectively. Even for $\langle a|\hat{O}_\mathrm{rand}|a \rangle$, these two values are going to be likely, although in many cases the pair $(j^\star,j^\star+1)$ will not be coupled directly by the SDRG procedure, but rather at a higher level, resulting in some intermediate value. 
However, in the ergodic regime --- when SDRG breaks down --- local thermalization implies that the local state of any pair of spins will be a uniform (at $ T=\infty$) mixture of the four possible above-mentioned states, resulting in a thermal average of zero for both observables.

\begin{figure}
\includegraphics[width=0.9\columnwidth]{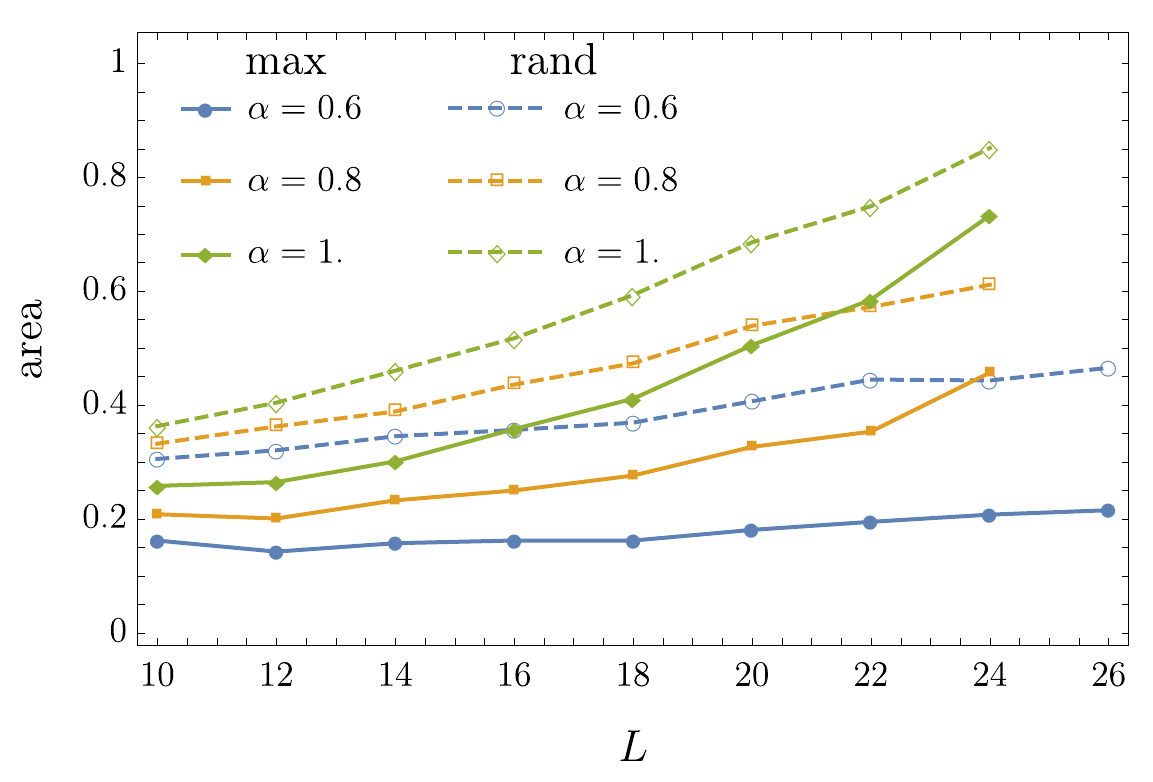}
\caption{Fraction of eigenstates with a value of $ \langle \hat{O}_\mathrm{max/rand} \rangle $ between $ -1/8 $ and $ 1/8 $, for $ \alpha \in \{0.6, 0.8, 1.0\} $ and $ S_0 = 0 $. ETH predicts this fraction to become 1 in the infinite-size limit.}
\label{Fig:ETH-areas}
\end{figure}

The distributions of the expectation values of $\hat{O}_\mathrm{max/rand}$ over eigenstates at system size $ L = 20 $ are shown in Fig.~\ref{Fig:ETH-heatmaps}.
It is clear that the system is perfectly compliant with the ETH at sufficiently high values of $\alpha$, whereas at smaller values of $\alpha$ the behavior consistent with the eigenstates being close to tree SDRG states. This phenomenology, which we interpret as a finite-size crossover between ergodic and nonergodic structure of the system's eigenstates, is compatible with the observed behavior for the level statistics (cf.\ Sec.~\ref{Sec:ShortScales_1}).

In order to validate our interpretation, we characterize the finite-size flow to ergodicity by looking at the percentage of eigenstates whose corresponding values of $\hat{O}_\mathrm{max/rand}$ falls within some fixed window centered at zero.
Fig.~\ref{Fig:ETH-areas} confirms that the ``ergodic fraction'' of infinite-$T$ eigenstates is increasing with $L$ for both $\hat{O}_\mathrm{max}$ and $\hat{O}_\mathrm{rand}$, though much more slowly for strong disorder. Crucially, at disorder $\alpha=0.6$, ETH is still strongly violated, which is consistent with the non-ergodic behavior observed in level statistics above. 


\subsection{Entanglement entropy}

Another witness of the non-ergodic behavior can be found in the scaling of the half-chain entanglement entropy with the system size, which is known to obey an area law for MBL systems, and a volume law for ergodic ones. More precisely, in a system that thermalizes, generic eigenstates are expected to be similar to random states; their entanglement entropy equals thermodynamic one, yielding for the states in the middle of the band: $S_{\rm ent}(L/2) = L/2 + o(L)$, when measured in bits~\cite{page1993information,amico2008entanglement}.

\begin{figure}
\includegraphics[width=0.9\columnwidth]{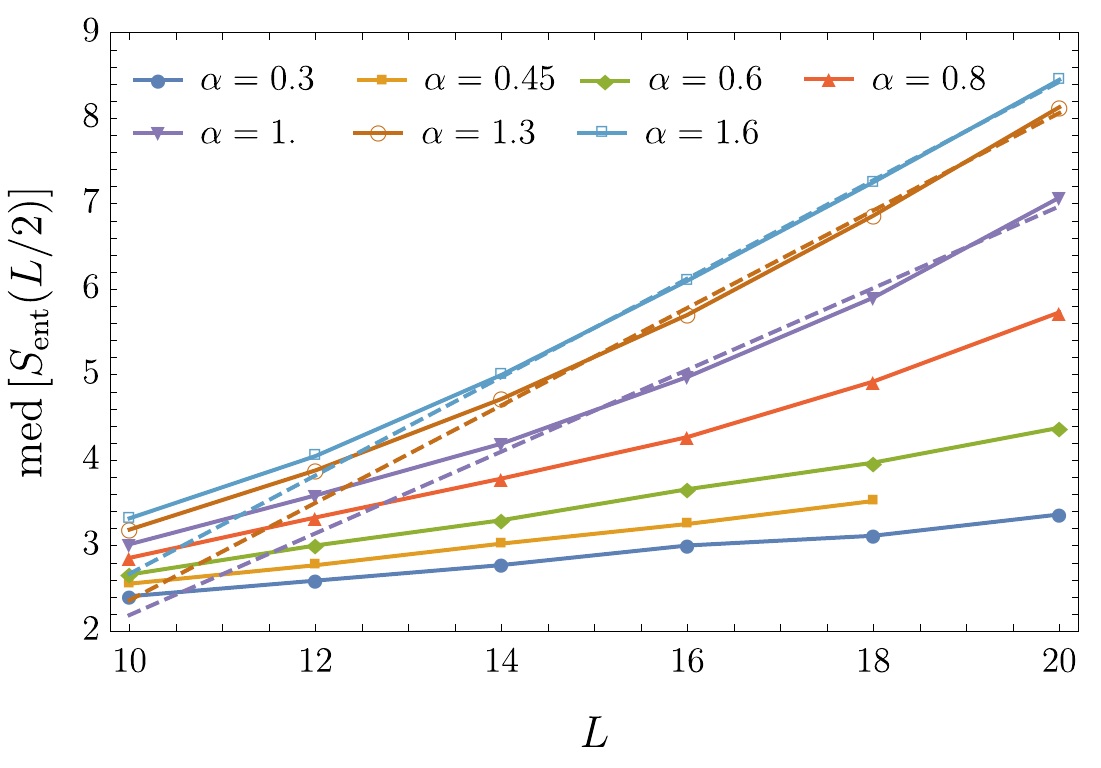}
\caption{A median value of the  half-chain entanglement entropy, $S_{\rm ent}(L/2)$, for the $S_0 = 0$ sector and $\alpha \in \{0.3,0.45,0.6,0.8,1.0,1.3,1.6\}$ (bottom to top). Linear fits, performed on the $ L \ge 14 $ points, are shown when their slope is close to that expected for an ergodic phase. Entanglement entropy for the higher disorder values is strongly sub-thermal.
 }
\label{Fig:median-ee}
\end{figure}

The numerical results are reported in Fig.~\ref{Fig:median-ee}. The median entanglement entropy of the infinite-temperature eigenstates exhibits linear scaling for all considered values of $\alpha$, but the linear coefficient observed at $L\leq 20$ deviates substantially from the ergodic prediction at strong disorder (although significant curvature is present). This is once more consistent with the results of Subsections~\ref{Sec:ShortScales_1} and \ref{Sec:ShortScales_ETH}.

To summarize the results of this Section, ED data show a clear trend towards ETH for moderate to weak disorder (i.e.\ $\alpha\gtrsim 0.6$) while indicating a novel non-ergodic regime  for the case of strong disorder. To determine behavior of the system in the thermodynamic limit, we have to resort to a completely different approach, presented in the next Section, which surpasses ED. 

\section{Resonance counting: from a single tree to a forest}

\label{Sec:Resonances}

As we showed in the previous Section, at strong disorder finite-size random Heisenberg chains exhibit a non-ergodic regime, in which their eigenstates are well-approximated by tree states. 
Here, to determine the eventual fate of these systems in the thermodynamic limit, $L\to\infty$, we develop an approach to analyze resonances between different tree states. We are able to capture long-range, multi-spin processes, which are beyond the conventional SDRG. We obtain the asymptotic behavior of the resonance number, and their spatial structure. 
We will find that the resonance density grows for all studied disorder strength, leading to an eventual delocalization at very large length scales, which we estimate. Beyond this length scale, the system presumably becomes ergodic. 

Given a tree state  generated by the SDRG, we can  construct a complete basis, Eq.~(\ref{eq:SDRGbasis}), in the Hilbert space (with the total spin of the system fixed) by allowing the values of the block spins identified by  SDRG to take all possible values consistent with the rules of angular momentum addition. The Hamiltonian~(\ref{Eq:Ham}), written in this basis, will then connect the initial SDRG state to a certain number of other tree states. We will consider the eigenvalue problem in this basis. The localization in this problem corresponds to true eigenstates being close to the tree states; in contrast, delocalization signals breakdown of SDRG approximation, suggesting ergodicity. 
The criteria for delocalization will be studied below.

\subsection{Connectivity of the hopping problem}
\label{Sec:Resonances_1}

\begin{figure}
\includegraphics[width=230pt]{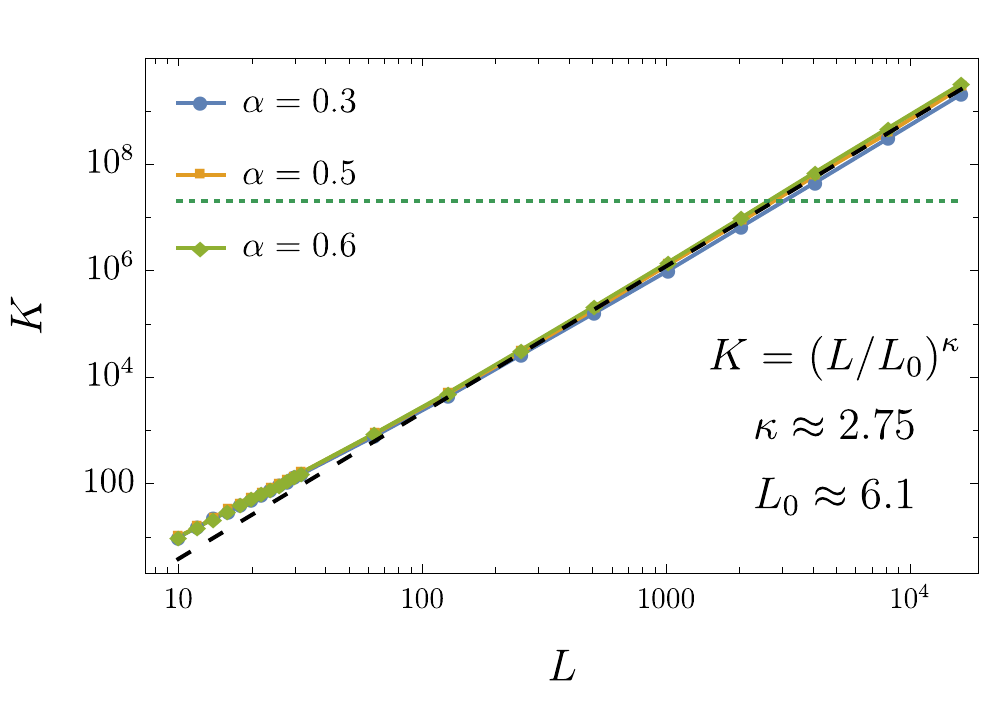}
\caption{Typical connectivity of the Heisenberg Hamiltonian in the tree basis obtained from SDRG, as a function of system size, for different values of disorder $\alpha=0.3,\, 0.5,\, 0.6,\, 0.8\,,1,\, 1.2$ (see legend).
The dashed line represents the fit $K=(L/L_0)^\kappa$. The extracted value of $\kappa=2.75$ is in good agreement with the analytic result (\ref{eq:Kkappa}), $\kappa=\ln3/\ln(3/2)\approx2.71$. 
The horizontal dotted line shows the maximal sampling rate ($2\times 10^7$ matrix elements) used in the search of resonances (see Secs.~\ref{Sec:Resonances_2} and \ref{Sec:Resonances_3}). 
}
\label{Fig:KTotal}
\end{figure}

First, we investigate the connectivity of this eigenvalue problem. That is, we analyze how many matrix elements of the Hamiltonian between a given tree state and other ones are non-zero. The $SU(2)$ symmetry of the model imposes stringent constraints on the matrix elements of the Hamiltonian~\cite{Protopopov2017}. Specifically, let us consider one of the terms in the Hamiltonian, $J_i {\bf S}_{i}{\bf S}_{i+1}$. It can be shown that the action of such an operator on a tree state can only affect the block spins that lie on the path in the tree connecting spins $i$ and $i+1$, see Fig.~\ref{Fig:schematic}(c).

Moreover, each of those block spins on the path, if affected, can only change by $0$ or $\pm 1$.  It then follows that the number of states connected to a given one by the operator  $J_i {\bf S}_{i}{\bf S}_{i+1}$ is given by:
\begin{equation}
K_{i, i+1}\simeq 3^{l_{i, i+1}}
\label{Kbasic}
\end{equation}
where $l_{i, i+1}$ is the length of the path in the tree connecting physical spins $i$ and $i+1$. The factor 3 arises from the selection rules: the operator can change the value of the representation at a node by $\Delta S=-1,0,+1$. The $\simeq$ sign  is due to the constraint that the new values of block spins in the tree  must still be consistent with the rules of angular momemtum addition (in particular, they cannot be negative). Sufficiently far from the bottom of the tree, the typical values of block spins are large and the latter constraint can only influence the prefactor in Eq.~(\ref{Kbasic}). Taking into account that the Hamiltonian~(\ref{Eq:Ham}) is just a sum of local terms of the form discussed above, we conclude that the total connectivity in the Hilbert space induced by the Hamiltonian (\ref{Eq:Ham}) is:
\begin{equation}
K\simeq \sum_{i=1}^L 3^{l_{i,i+1}}.
\label{eq:Ksum}
\end{equation}
We are now left with the task of computing the distribution $P(l)$ of these lengths $l_{i,i+1}$, for  the SDRG trees. The  SDRG fuses spins that are most strongly coupled. Neglecting the correlations between the (renormalized) couplings  at any step of SDRG as well as the dependence of those on the couplings at \emph{earlier} stages of SDRG we can assume that  the pair of spins to be fused is just randomly chosen among all possibilities (with the only requirement that the fusing spins are nearest neighbors so that locality is respected).  In such an ensemble of \emph{maximally random} SDRG trees the distribution $P(l)$ can be computed analytically. As we show in Appendix~\ref{App:distance_distribution}, it turns out that $P(l)$ falls down exponentially with $l$, and in the limit $L\to\infty$ it becomes:
\begin{equation}
P(l)=\frac{3}{4} \left(\frac{2}{3}\right)^l
\label{eq:PL}
\end{equation}
(the normalization is the correct one considering $l\geq 2$, so $\sum_{l\geq 2}P(l)=1$). With this distribution $P(l)$, the sum (\ref{eq:Ksum}) is dominated by the maximum $l_M$ over the $L$ terms. To the leading order in $L$, the value of $l_M$ can be estimated from the condition $LP(l_M) \sim 1$. This follows from the distribution of the largest of $L$ random variables with the distribution (\ref{eq:PL}), which is given by $P(l_M=x)=\frac{3L}{4} \left( \frac 23  \right)^x \left(1- \left(\frac 23\right)^{x-2} \right)^{L-1}$.

This yields 
\begin{equation}
l_M=\max_{i=1,\dotsc,L} l_{i,i+1}\sim\frac{\ln L}{\ln(3/2)}.
\end{equation}
Plugging back into $K\sim 3^{l_M}$ we find
\begin{equation}
K\sim L^\kappa,
\label{eq:Kkappa}
\end{equation}
with $\kappa=\ln{3}/\ln(3/2)\simeq 2.71$. The power-law scaling~(\ref{eq:Kkappa}) and the value of the exponent $\kappa$ are in a good agreement with the numerical simulations of the SDRG trees (with the full set of SDRG rules taken into account), see Fig.~\ref{Fig:KTotal}.

\begin{figure}
\includegraphics[width=230pt]{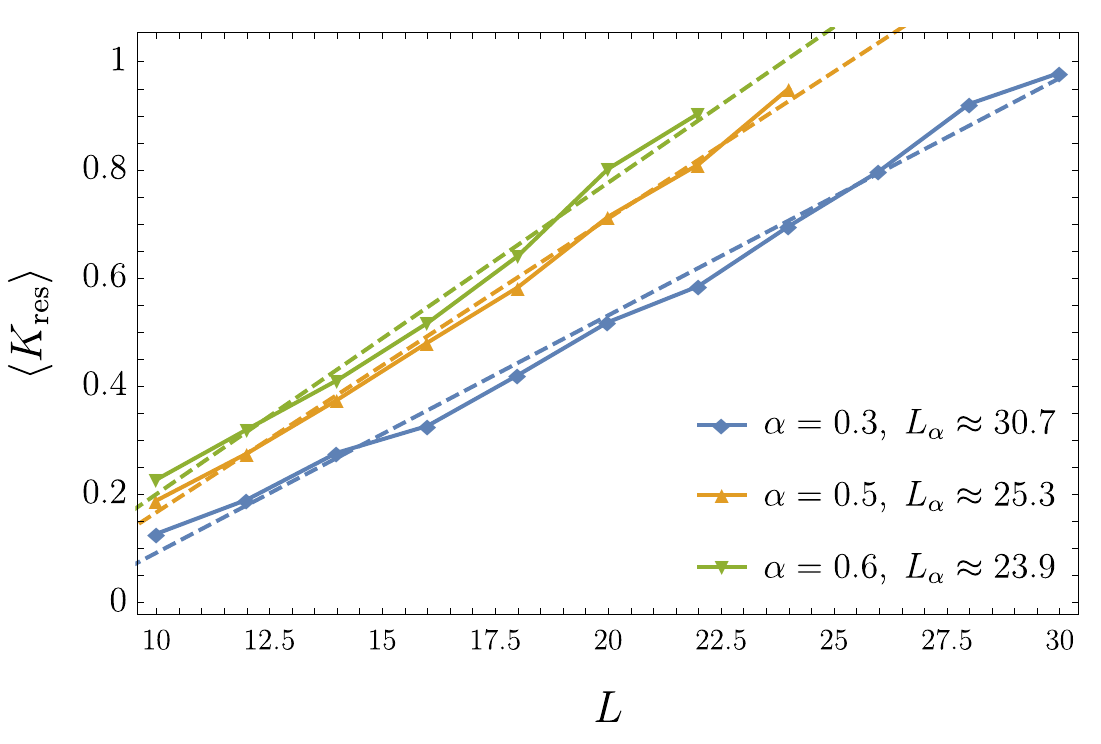}
\caption{ 
The average number of resonant neighbors for an SDRG tree state in short systems.
The solid lines of different colour correspond to different values of the exponent $\alpha$.  Dashed lines represent linear fits,  $\langle K_{\rm res}\rangle =a L-b$.
For each value of disorder the legend also indicates the scale $L_1(\alpha)$ defined by   $\langle K_{\rm res}\rangle=1$ (see main text). 
 }
\label{Fig:ResNumberShort}
\end{figure}

\subsection{Local resonances}
 \label{Sec:Resonances_2}
 
Our next goal is to find resonances among the $K\propto L^{\kappa}$ ``hopping" processes generated by the Heisenberg Hamiltonian for a given tree state. We will first focus on investigating relatively small system sizes, $L\lesssim 30$, comparable to those accessible by exact diagonalization.  

To study the number of resonances, we first use SDRG procedure to generate a random tree state $|\Psi^0_{\rm RG}\rangle$ and identify resonant neighbors for that state -- that is, the ones for which the ratio of matrix element connecting them to $|\Psi^0_{\rm RG}\rangle$ and the energy difference is larger than one.  These resonances invalidate the perturbative expansion around the ``infinite disorder'' eigenstates (the SDRG states). Their proliferation signals the instability of tree states, strongly suggesting that ergodicity is restored. The SDRG is essentially a local optimization procedure that aims to construct basis states free of such resonances. Based on the results presented above, we expect that, at strong disorder and in relatively short systems, these resonances should be few in number, because SDRG is accurate.

 The average number of resonant neighbours, $\langle K_{\rm res}\rangle$, of an SDRG tree state is shown in Fig.~\ref{Fig:ResNumberShort}. We observe that for relatively small systems discussed here, $\langle K_{\rm res}\rangle$  scales linearly with the system size $L$. As expected, the slope of this linear growth becomes smaller for stronger disorder.  

The condition $\langle K_{\rm res}\rangle=1$ defines an important (disorder-dependent) lengthscale in the problem, $L_1(\alpha)$, at which resonances start appearing.  Naively, this lengthscale plays the same role as the lengthscale $L_W$ introduced in Sec. \ref{Sec:RGandTreeStates_1} to characterise the resonances in a random-field XXZ chain.  
We found that the scale $L_1(\alpha)$ grows at stronger disorder, crudely following a power-law dependence, $L_1(\alpha)\propto \alpha^{-0.4}$.

Fig.~\ref{Fig:ResNumberShort} shows that  at relatively strong disorder values, $\alpha\leq 0.6$,  the average  number of resonant neighbors 
for an SDRG tree state is 1 or less  for all system sizes available in ED. This agrees with the observation that such chains display a non-ergodic behavior in all of the ED studies of Sec. \ref{Sec:ShortScales}, with eigenstates being well-approximated by the tree states.

In particular, the low number of resonances is in agreement with the slow growth of $N_E$ (the participation ratio of eigenstates in the tree basis) found in Sec.  \ref{Sec:ShortScales_2}.  Drawing parallels to the conventional MBL systems, it is tempting to identify the length scale $\tilde{L}_1(\alpha)$ that controls the exponential growth of $N_E$ with the system size (see Sec.\ \ref{Sec:ShortScales_2}) with $L_1(\alpha)$. However, the comparison of the values of $L_1(\alpha)$ and $\tilde{L}_1(\alpha)$ reveals that the latter is several times shorter. We attribute this difference to the effect of the second-order perturbative corrections that contribute to the spreading of the exact eigenstate $|E\rangle$ over SDRG tree state.  Such higher order perturbative corrections lie beyond the first order resonance counting that underlies the scale $L_1(\alpha)$. The perturbative corrections are expected to be more significant at weak disorder;  in accordance with this intuition, we found a more significant difference between $L_1(\alpha)$ and $\tilde{L}_1(\alpha)$ for such disorder strengths.

\subsection{Longer systems and the proliferation of resonances }
\label{Sec:Resonances_3}

Does linear scaling of $K_{\rm res}$ with the system size discussed in the previous Subsection persist  in the thermodynamic limit? Such behavior would closely resemble that of the strongly disordered XXZ model. It would imply that the resonant neighbours can be attributed to the existence of local subsystems with resonating levels which, if sufficiently separated in space, would remain isolated and would not cross-talk (in the sense that there is no significant entanglement in the eigenstates between such ``local" resonances). If true, this would be a strong argument in favour of the SDRG  tree states surviving in an infinitely long system, up to corrections due to local, isolated resonances. We now perform a detailed analysis of resonances in large systems, up to $L\sim 2 \times 10^3$, and find that Heisenberg chains actually behave qualitatively differently compared to the plain-vanilla MBL systems: the number of resonances grows {\it faster} than linear with the system size.


\subsubsection{Number of resonances and their structure}
\label{Sec:Resonances_3_1}
\begin{figure}
\includegraphics[width=230pt]{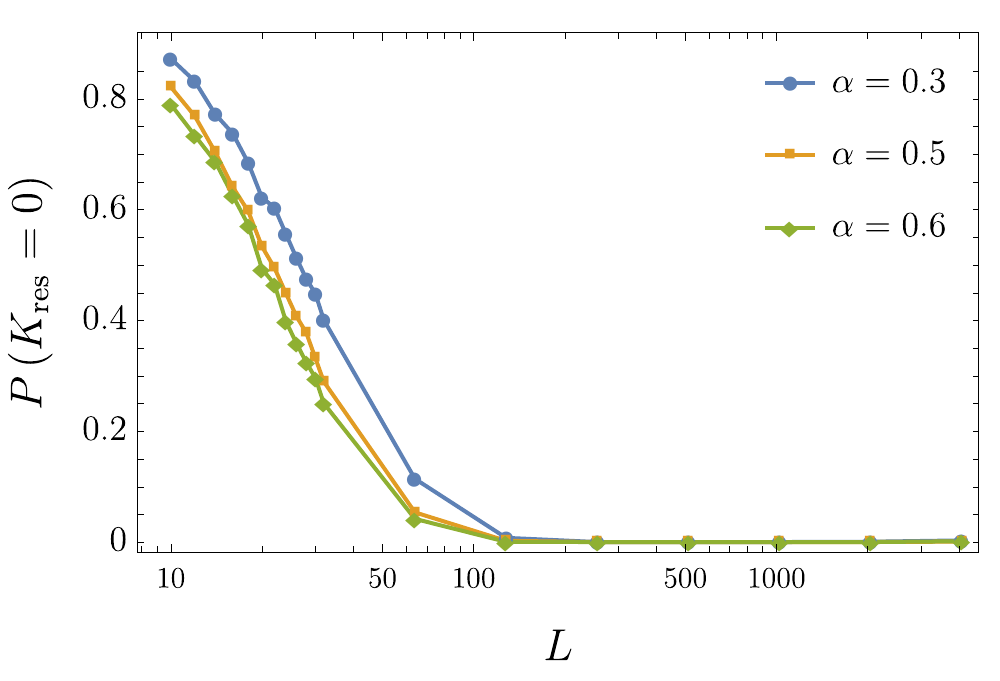}

\hspace{0.5cm}
\includegraphics[width=230pt]{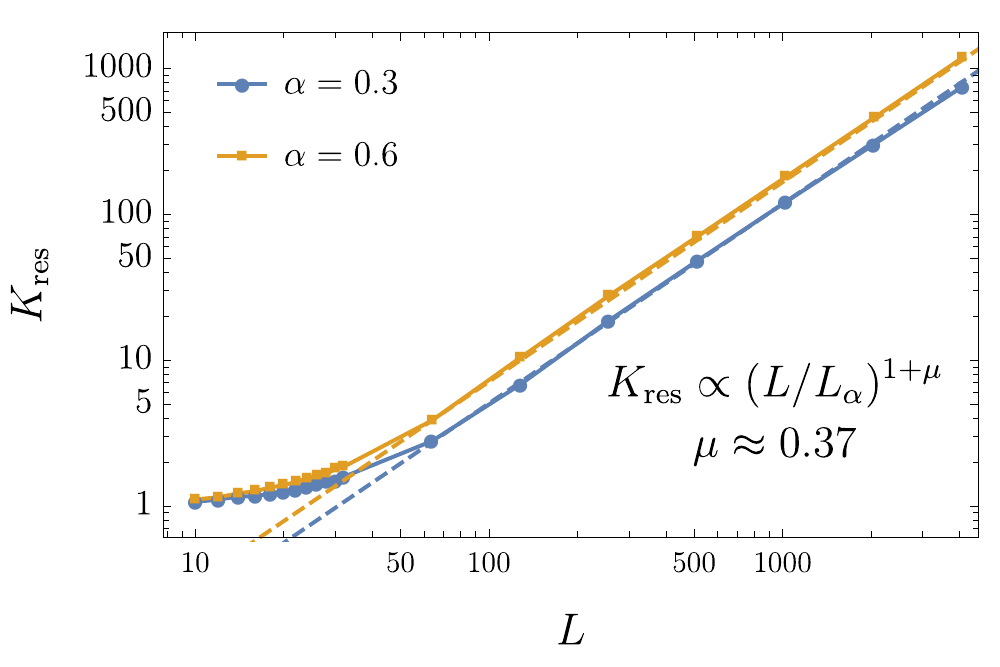}
\caption{
(Top) The probability $P(K_{\rm res}=0)$ for an SDRG tree states to have no resonant neighbors, as a function of the system size, and for different values of disorder (see legend). 
Note the logarithmic scale along the horizontal axis. While short systems are essentially free of resonances even for a relatively weak disorder,  the probability $P(K_{\rm res}=0)$ becomes vanishingly small in long systems. 
(Bottom) A typical number of resonant neighbours for an SDRG tree state in long systems, $L \leq 2^{12}$. The  three solid lines of different colour correspond to the two different values of the exponent $\alpha=0.3,\, 0.6$. The dashed lines show the power-law fits, $K_{\rm res} =\left[L/L_1(\alpha)\right]^{1+\mu}$. The scale $L_1(\alpha)$ where the average number of resonant neighbors for an SDRG tree state equals 1 was defined in the previous Subsection.   
 }
\label{Fig:ResNumberLong}
\end{figure}

 The probability for an SDRG tree state to have no resonant neighbours vanishes in sufficiently long systems (see top panel in Fig.~\ref{Fig:ResNumberLong}). Then, a typical tree state has a large number of resonances attached to it.  
A bottom panel in Fig~\ref{Fig:ResNumberLong}  shows (in log-log scale) the dependence of the {\it typical} number of resonant neighbors (defined as $e^{\langle \ln K_{\rm res} \rangle}$) for an SDRG tree state.  The dashed lines represent power-law fits:
\begin{equation}
K_{\rm res}\propto \left(\frac{L}{L_1(\alpha)}\right)^{1+\mu}
\end{equation}
with the scale $L_1(\alpha)$ determined from the the short-scale behavior of $K_{\rm res}$, see Sec. \ref{Sec:Resonances_1}.  The ``anomalous'' exponent $\mu$ is approximately disorder-independent, $\mu\approx 0.38$. The details of the numerical procedure employed to find and characterize  resonances are given in Appendix \ref{App:ResonanceSearch}. 
\begin{figure}
\includegraphics[width=230pt]{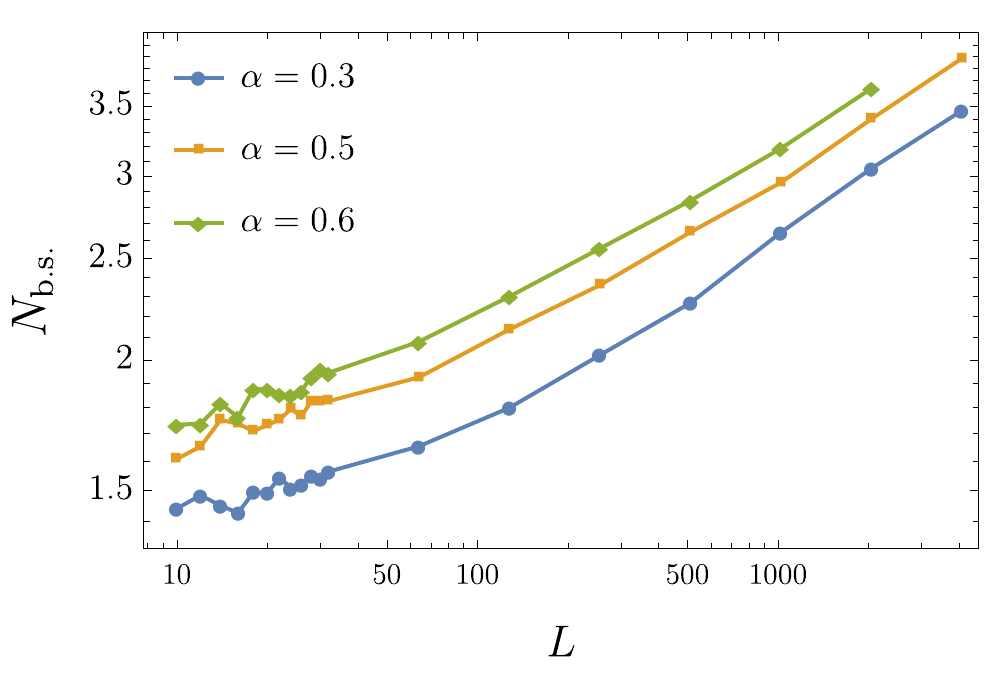}
\includegraphics[width=230pt]{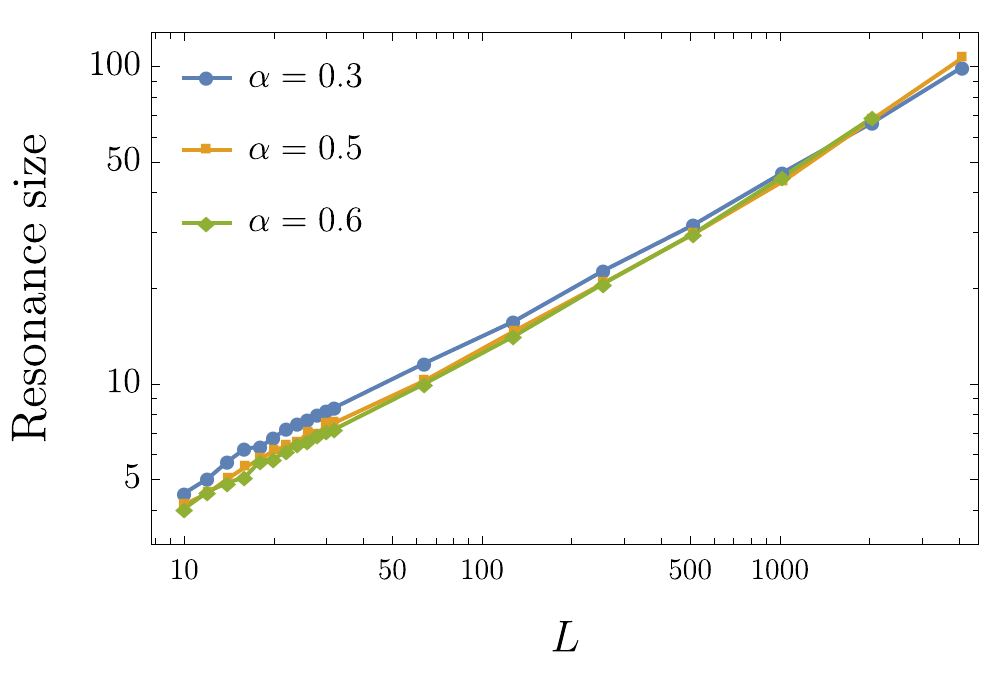}
\caption{ (Top) Average number of the block spins changing their  value in a single resonant transition.
(Bottom) Typical number of adjacent physical spins involved in a resonance (see main text).  
}
\label{Fig:ResStructure}
\end{figure}

\begin{figure}
\includegraphics[width=230pt]{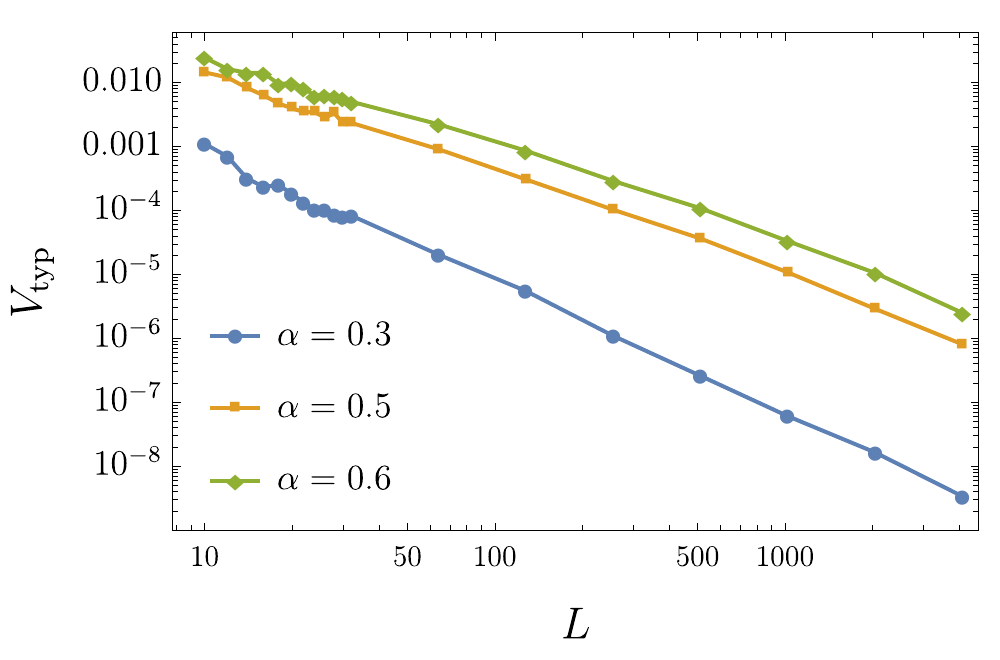}
\caption{
Characteristic energy scale for typical resonances, as a function of system size, shown for different disorder strengths. 
 }
\label{Fig:ResEnergy}
\end{figure}

The power-law scaling of the number of resonant neighbors, $K_{\rm res }\propto L^{1+\mu}$, implies that, in stark contrast to the random-field XXZ model,  the {\it density}  of the resonating degrees of freedom  grows with the system size. 
Accordingly, at least some of the resonant transitions must  originate not from the rearrangement of a few local spins, but rather involve a growing number of spins. To support this conclusion, we analyze the structure of typical resonances.  
The top panel of Fig.~\ref{Fig:ResStructure} shows the average number of block spins, $N_{\rm bs}$, that are changed in  the course of a resonant transition. We observe that $N_{\rm bs}$ grows (albeit rather slowly) with the length of the chain.   The decrease of $N_{\rm bs}$ with increasing disorder can be understood as follows: at weak disorder the possibility to change a block spin in a resonant manner is often accompanied by an ``instability'' (with respect to resonances) of the block spins higher up in the hierarchy.
The more complicated (involving flipping of more than one block spin) resonant neighbors appear and contribute to the increase of the average $N_{\rm bs}$. On the other hand, at strong disorder an ``instability'' of a single block spin is more likely to remain ``localized'' and  and not to ``propagate'' upwards in the tree. 

The size of a typical resonance in {\it real space} also grows as the system size is increased, see the bottom panel of Fig.~\ref{Fig:ResStructure}. An elementary physical spin is affected by a resonant transition if at least one of its descendant block spins changes its state. The physical size of a resonance is then defined as the total number of the elementary spins involved in it. Essentially, it is the level (as counted from the bottom of the tree) of the highest block spin affected by the resonance that sets the size of the resonance. 


We observe that at moderate system sizes the typical spatial size of a resonance at strong disorder exceeds that at weak disorder. This is in accord with our intuition: at strong disorder a large number of spins need to rearrange collectively in order for a transition to be resonant.  In terms of the SDRG, this means that many SDRG steps can be performed before the resonances start to play any role. On the other hand, in  sufficiently long systems we see the opposite tendency: weakly disordered chains typically exhibit resonances of larger  size.   This is the manifestation of the propagation of an ``instability'' of block spins upwards in the tree, cf. discussion above.

The growing lengthscale characterizing the resonances comes together with a decreasing energy scale. The latter is given by a typical matrix element for a resonant transition,  $V_{\rm typ}$. Its system size dependence is shown in Fig.~\ref{Fig:ResEnergy}.  In the following Subsection, we will use $V_{\rm typ}$ to estimate the energy scale associated with the crossover to ergodicity.

\subsubsection{Breakdown of SDRG and delocalization}
\label{Sec:Resonances_3_2} 
The results  presented in  the previous Subsection (most importantly, the power-law growth of the resonance density) strongly suggest that even in the strongly disordered chains with $\alpha\leq 0.6$, where ED studies of  Sec. \ref{Sec:ShortScales} reveal little (if any) signs of ergodicity, the resonant transitions missed by SDRG eventually proliferate. In this Subsection, we estimate the corresponding thermalization scale $L_{\rm erg}(\alpha)$.


Given an SDRG tree state  and a set of resonant transitions associated with it, one can identify a set  of block spins that can be changed via at least one resonant process. We refer to those block spins as resonant, or unstable ones. For a chain of $L$  spins there are $2L-1$ nodes in the SDRG tree ($L$ of them are leaves corresponding to the physical spins).  At each stage of the SDRG procedure some number $L_{\rm RG}$ of the block spins  play the role of the physical spins of the system.  
For example, in the initial state of SDRG $L_{RG}=L$ and the SDRG spins are just the physical ones. The final stage of SDRG corresponds to $L_{\rm RG}=1$, and the top node of the SDRG tree being the only remaining spin. The ratio $L/L_{RG}$ is nothing but the average size of the spin clusters in the system.  

At any given moment in the course of the SDRG, only the unstable block spins that are among the $L_{\rm RG}$ spins currently comprising the system, are relevant for potential delocalization. The others either have not yet formed, or have already been decimated by SDRG; theyare not expected to contribute directly to the  physics at the current energy scale. It is thus natural to ask how  the number and density of the resonant block spins evolve in the course of the SDRG.


\begin{figure}
\includegraphics[width=210pt]{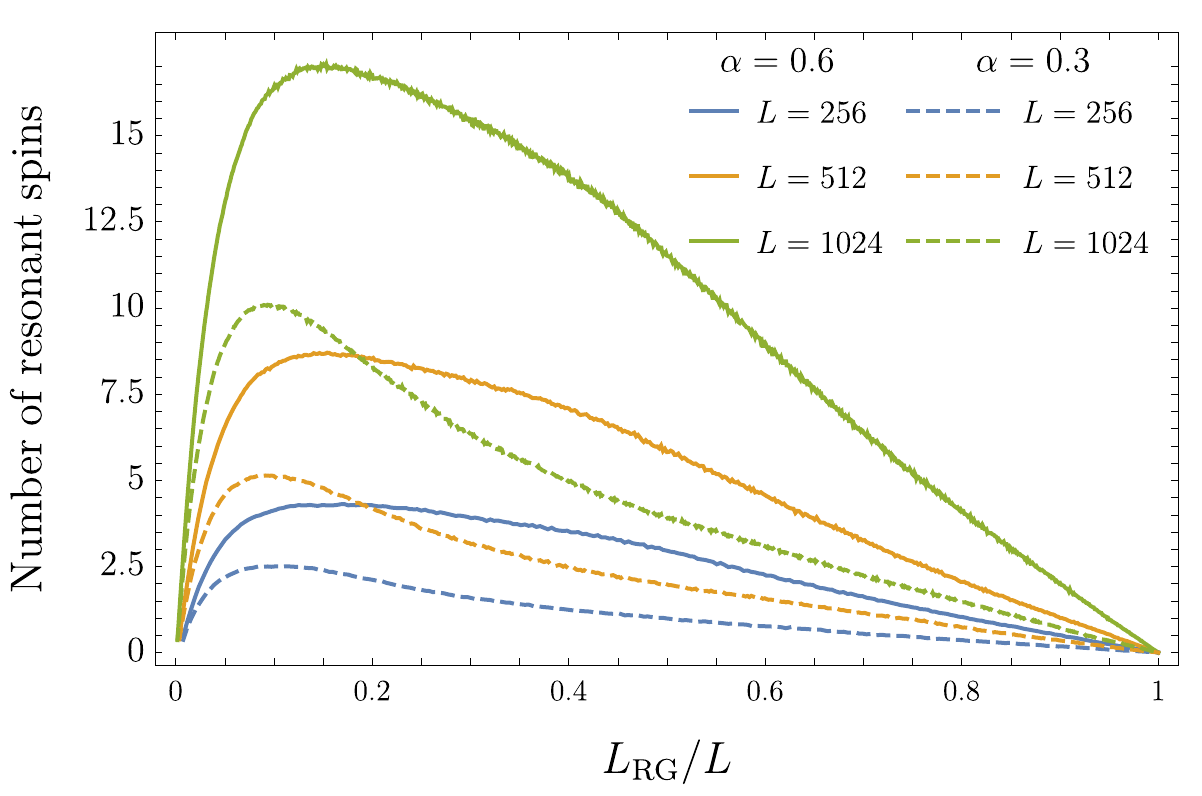}
\includegraphics[width=210pt]{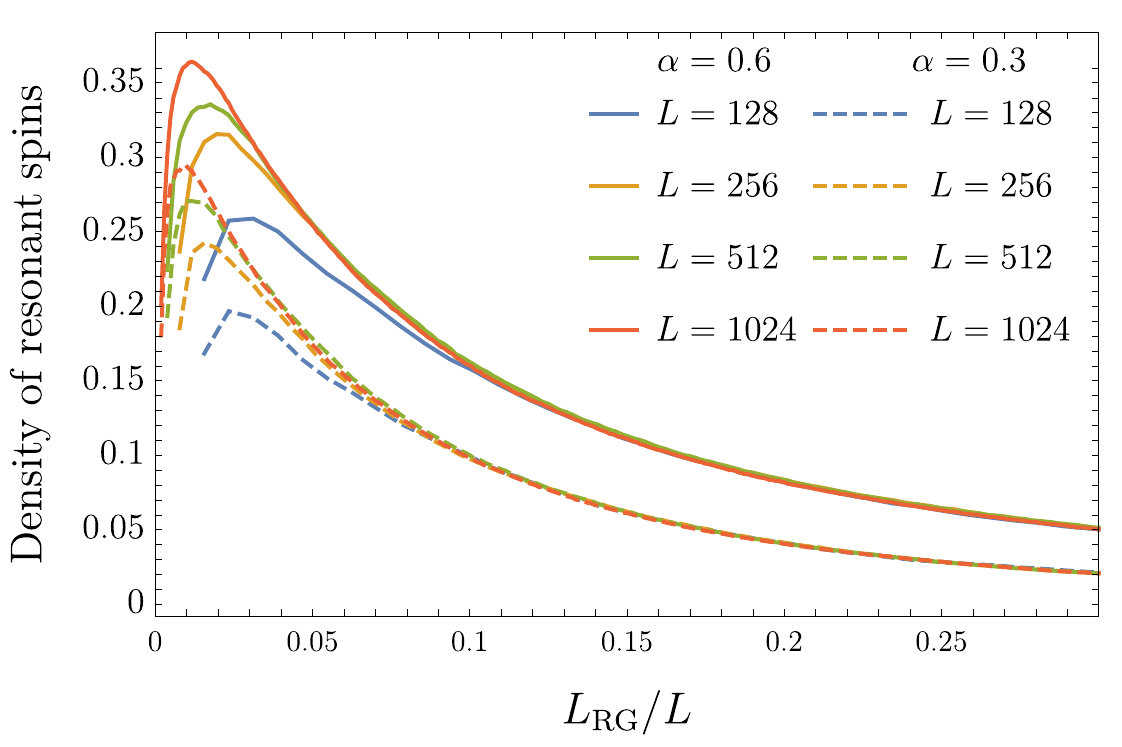}
\caption{Evolution of the number (top) and the density (bottom)  of the resonant blocks spins in the course of SDRG. The horizontal axes shows the ratio of the current system size to the initial length of the system. 
 }
\label{Fig:RGResSpins}
\end{figure}

\begin{figure}
\includegraphics[width=230pt]{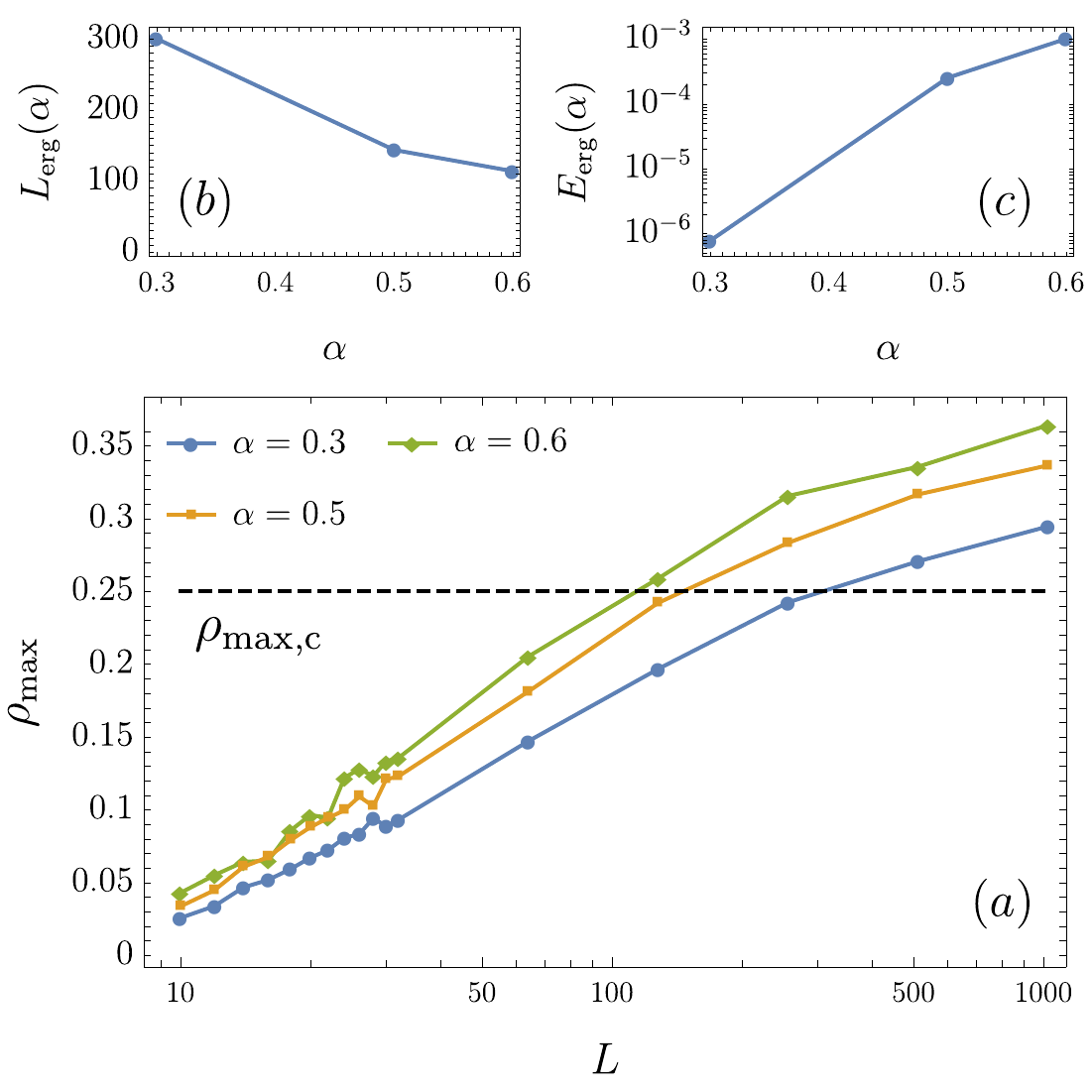}
\caption{(a) Maximum density $\rho_{\rm max}$ of the resonant spins developed in the course of SDRG  (cf.  Fig. \ref{Fig:RGResSpins}) as a function of system size. The  condition $\rho_{\rm max}\sim \rho_{\rm max, c}$ defines the ergodization  length scale $L_{\rm erg}(\alpha)$ and the corresponding energy scale $E_{\rm erg}(\alpha)=V_{\rm typ}\left[L_{\rm erg}(\alpha)\right]$. 
(b) and (c) Estimates for the ergodization length and energy scales obtained by fixing the critical density to  $\rho_{\rm max, c}=0.25$. 
 }
\label{Fig:RGResSpinsMax}
\end{figure}

This is illustrated in Fig.~\ref{Fig:RGResSpins} that shows the dependence of the number (top panel) and the density (bottom panel) of unstable block spins for two different values of $\alpha$, and several values of the physical chain length $L$. These quantities are plotted as a function of the running RG length $L_{\rm RG}$, normalized by $L$. 
For not  too small $L_{RG}/L$, we observe that for a fixed disorder strength, the density of resonant spins exhibits a universal ($L$-independent) behavior, $\rho_{\rm res}(L_{\rm RG}/L, L)=\rho_{\rm res}(L_{\rm RG}/L)$.   The density $\rho_{\rm res}(L_{\rm RG}/L)$ is higher at weaker disorder and also  grows in the course of SDRG. In contrast, at small $L/L_{\rm RG}$ (corresponding to the final stages of SDRG in a finite chain) a rather pronounced dependence on $L$ is observed.  
{

Next, let us denote by $\rho_{\rm max}\equiv\rho_{\rm max}(\alpha, L)$ the maximum density of unstable spins developed during SDRG process. Small $\rho_{\rm max}$ means that $\rho_{\rm res}$ remains small at all steps of SDRG. We then expect the resonances to be of little importance for our system. 
   On the contrary, $\rho_{\rm max}\sim 1$, indicates that at some stage, the SDRG inevitably runs into a state where almost  all  spins participate in resonances. Then, the basic assumptions of SDRG are violated and we expect it to break down -- this means that block-spins are no longer well-defined, and start resonating. Presumable, this signals the onset of ergodicity.
   
It is natural to assume that there exists a critical value $\rho_{\rm max, c}<1$ at which the crossover between non-ergodic (SDRG valid) and ergodic (breakdown of SDRG) regimes occurs. We can then identify the length of the system for which $\rho_{\rm max}=\rho_{\rm max, c}$, as the ergodicity scale: 
 \begin{equation}
 \rho_{\rm max}\left[\alpha, L_{\rm erg}(\alpha)\right]=\rho_{max, c}.
   \end{equation}
The scale $L_{\rm erg}(\alpha)$ along with  the typical matrix element for resonant  transitions $V_{\rm typ}$ gives an estimate for the ergodicity time and energy scales: 
\begin{equation}
\tau_{\rm erg}=E_{\rm erg}^{-1}, \;\; E_{\rm erg}=V_{\rm typ}(L_{\rm erg}).
\end{equation} 

Figure~\ref{Fig:RGResSpinsMax} shows the dependence of $\rho_{\rm max}(L)$ for different disorder strengths. Estimating $L_{\rm erg}(\alpha)$ requires fixing the critical density $\rho_{\rm max, c}$. While we have no  general theory for $\rho_{\rm max, c}$, we observe (see Fig.~\ref{Fig:RGResSpinsMax}) that the 
the value $\rho_{\rm max, c}\in [0.2, 0.25]$ (similar to the critical density of resonances in the random XXZ model) results in an estimate $50\lesssim L_{\rm erg}(\alpha=0.6)\lesssim 100$ that is roughly consistent with the intuition developed in ED studies of Sec. \ref{Sec:ShortScales_1}, $L_{\rm erg}(\alpha=0.6)\sim 2 L^*(\alpha=0.6)\sim 50$.  
Thus, for the purpose of an estimate, we choose  $\rho_{\rm max, c}=0.25$. The resulting values for the lengthscale and energy scales at which thermalization starts to occur, are shown in Fig.~\ref{Fig:RGResSpinsMax}. 

It is evident that at strong disorder, $\alpha=0.3$, resonances start to proliferate only at very large lengthscales $L_{\rm erg}\approx 300$, and, moreover, the corresponding time scales are extremely long. Such time scales are beyond the limitations of the synthetic platforms, where ergodicity and its breakdown are actively investigated (see Ref.~\cite{AbaninReview} for a review). Thus, in experiments, strongly disordered, $SU(2)$ symmetric systems are expected display the novel non-ergodic regime described above. 


Systems of size $L\gg L_{\rm erg}$ will be slowly thermalizing, and will presumably display slow diffusive transport at low frequencies. An interesting open question concerns the eventual fate of the integrals of motion obtained in the first steps of the SDRG (when the typical cluster size is much smaller than $L_{\rm erg}$). Such nearly-conserved operators arise due to strongly coupled clusters spins, and therefore destroying them would typically involve a relaxation process with a large energy scale $\Delta E$. In very large systems, slow thermalizing processes will eventually destroy the conservation of these operators. However, since thermalization processes typically occur on a much smaller energy scale, $E_{\rm erg}\ll \Delta E$, we expect that the decay time of such operators will be parametrically large in $\Delta E/E_{\rm erg}$. An instructive example is that of a narrow-bandwidth thermal bath with energy scale $E_0$; there, the relaxation of excitations with energy $\omega\gg E_0$ is exponentially slow in $\omega/E_0$~\cite{Abanin15}. We expect that the integrals of motion obtained within SDRG before its breakdown, will be similarly long-lived (but we leave a detailed investigation of this issue for the follow-up work). Thus, we propose a picture that the dynamical properties of the strongly disordered Heisenberg chains, will be captured by SDRG at frequencies~$\omega\gtrsim E_{\rm erg}$ (in particular, they would have non-trivial noise properties, described in Ref.~\cite{Agarwal2015}). At lower frequencies, $\omega\lesssim E_{\rm erg}$, a crossover to a diffusive behavior is expected.

\section{Conclusions and outlook}

To sum up, the goal of this paper was to investigate the effects of continuous non-Abelian symmetries on dynamical properties of disordered systems. We have considered 
a concrete example of disordered Heisenberg spin chains, which are characterized by an $SU(2)$ symmetry. To describe the properties of this model, we combined stat-of-the-art exact diagonalization studies with a new approach that allows us to include long-range resonances into the strong-disorder renormalization group. 

We have found that in a broad range of disorder strengths and system sizes, Heisenberg chains exhibit a new kind of non-ergodic behavior. In this regime, the highly excited eigenstates have a scaling of entanglement entropy that is intermediate between the area-law characteristic of MBL states, and the volume-law found in thermalizing systems. This behavior stems from the tree tensor-network structure of the eigenstates obtained within SDRG. Simultaneously, in this regime the system exhibits a different kind of integrability, with integrals of motions having a varying degree of locality: some of them act on a small number of neighboring spins, while others act on larger and larger spin clusters. 

Further, we found that for weak disorder, the behavior crosses over from non-ergodic to ergodic as the system size is increased. For stronger disorder, all system sizes accessible numerically exhibited non-ergodic behavior. To address the eventual fate of the non-ergodic phase in this case, we have extended SDRG approach, characterizing resonances that endanger the stability of tree states. Our results strongly suggest eventual delocalization and ergodicity, albeit at very large system sizes; delocalization occurs via unconventional, multi-spin processes, which is yet another unique feature of disordered systems with non-Abelian symmetries. In the future work, we plan to describe the transition between non-ergodic and ergodic regimes as a function of system size. A promising starting point seems to be to formulate an effective model in terms of resonant degrees of freedom, with parameters extracted using methods described above.

Another interesting direction is to better understand dynamical signatures of the new non-ergodic regime uncovered here. One natural experiment would be to probe the dynamics of the most local integrals of motion (e.g.\ total spin of a pair of strongly coupled physical spins), and to observe that, for system sizes $L< L_{\rm erg}$ it is conserved to a good precision and for arbitrarily large times. Another interesting open question concerns spin transport in $N$-species, disordered Hubbard models~\cite{Bloch15,PrelovsekHubbard1}. In case of flavour $SU(N=2)$ symmetry, our work suggests that a sufficiently large system should show thermalizing behavior. Further work is required to establish the details of the dynamics (e.g.\ diffusion vs.\ subdiffusion).

More broadly, this work sets the stage for future discovery of new non-ergodic regimes and true dynamical phases that survive in thermodynamic limit.  The approach introduced here can be naturally extended to other symmetry groups, for example $SU(N)$ spins. We leave studies of such systems for future work. Even more generally, it would be interesting to investigate the stability of other tree tensor network structures with intermediate entanglement scaling, as possible good approximation of eigenstates in physical systems.

\acknowledgements We thank D. Huse, A. Mirlin, M. Serbyn, and R. Vasseur for illuminating discussions. We also thank K. Agarwal, W. W. Ho and I. Martin for previous collaboration on related topics. This research was supported by the Swiss National Science Foundation (IVP and DAA), by CINECA, ISCRA grant: project IsC66 EDNAS (RKP, TP, AS), and by Harvard-MIT CUA, AFOSR Quantum Simulation MURI, AFOSR-MURI: Photonic Quantum Matter (award FA95501610323) (ED).

\bibliography{mbl.bib}


\appendix
\section{SDRG for Heisenberg spin chains}
\label{App:SDRG}
The SDRG procedure for Heisenberg spin chains was formulated and discussed  comprehensively in a number of publications \cite{ma1979random,Dasgupta1980, Fisher1992,  Westerberg1997, Agarwal2015}. As is common for the RG studies, the aforementioned works focused on the flow of the system parameters under RG transformation and the consequences of this flow for the thermodynamic properties of the system. The interpretation of SDRG approach from the perspective of the many-body eigenstates  (including highly excited ones) was put forward in Ref. \cite{Pekker14}. In this Appendix we briefly review the SDRG protocol  for $SU(2)$-symmetric Heisenberg spin chains with the emphasis on this later aspect of the problem. We also discuss several subtle points of the procedure.

The SDRG protocol we design deals with the spin Hamiltonian of the form 
\begin{equation}
H=\sum_{i} H_i,  \qquad H_i = J_i {\bf S}_{i}{\cdot\bf S}_{i+1}.
\label{App:Eq:Ham}
\end{equation}
In the initial state of SDRG the spin operators ${\bf S}_i$ represent the elementary spins $1/2$ that constitute the system. The summation runs over nearest-neighbors links in a 1D lattice, $i=1, \ldots, L-1$. The SDRG procedure merges individual spins into clusters. Correspondingly, at later stages of SDRG  ${\bf S}_i$ represent the total angular momentum of clusters of elementary spins (block spins in the terminology of Sec. \ref{Sec:RGandTreeStates_2}). The corresponding quantum number $S_i$ 
can take arbitrary integer or half-integer values  limited from above by half of the size of the cluster.   

The eigenstates and eigenvalues of each of the ``link'' Hamiltonians $H_i$ are completely fixed by symmetry. Its spectrum consists of $n_{i}\equiv 2\min(S_i, S_i+1)+1$ levels 
with energies
\begin{equation}
E_{i, \tilde{S}_i}=\frac{J_i}{2}\left(|\tilde{{\bf S}}_i|^{2}-|{\bf S}_{i}|^2-|{\bf S}_{i+1}|^2\right) 
\label{App:Eq:Eis}
\end{equation}
where by $|{\bf S}|$ we denote the absolute value of the spin, $|{\bf S}|\equiv\sqrt {S(S+1)}$, and   $\tilde{S}_i=|S_i-S_{i+1}|,\, |S_i-S_{i+1}|+1\,, \ldots, S_i+S_{i+1}$ stands for the total spin of the cluster formed by the spins ${\bf S}_{i}$ and ${\bf S}_{i+1}$. 

Every link $i$ in the system is thus associated with a set of energy gaps in the ``link'' Hamiltonian $H_i$
\begin{equation}
\Delta_{i, \tilde{S}_i}^{\pm}=\left|E_{i, \tilde{S}_{i}}-E_{i, \tilde{S}_{i}\pm 1}\right|, \qquad 
|S_i-S_{i+1}|\leq\tilde{S}_i\leq S_i+S_{i+1}\,.
\label{App:Eq:Delta}
\end{equation}
The gaps $\Delta_{i, \tilde{S}_i}^{\pm}$ have the physical meaning of the precession frequencies for the 
vector ${\bf S}_{i}-{\bf S}_{i+1}$ in the state of the $i$-th link characterised by the total spin $\tilde{S}_{i}$. 

The SDRG procedure aims to eliminate from the system the fastest degrees of freedom. Therefore \cite{Agarwal2015}, it looks for the link  $i_0$ with maximal value of $\underset{\pm}{\operatorname\min\,}\Delta^{\pm}_{i, \tilde{S}_i}$ and approximates the the state of the link  $i_0$  by the one with definite total spin $\tilde{S}_i$. Thereby we eliminate from the consideration the rapidly oscillating  vector~${\bf S}_{i_0}-{\bf S}_{i_0+1}$.

Note a subtle point here: at any stage of SDRG each link in the system is generically characterized by a set of $n_i-1>1$ energy gaps and the judgement on  which link represents the strongest-coupled subsystem requires a {\it guess}  about the total  spin $\tilde{S}_i$ associated to each  link.  Our present situation is to be compared with the SDRG for the ground state of the Heisenberg spin chains  \cite{ Westerberg1997}  or SDRG for the highly excited states of less symmetric systems 
\cite{Pekker14}. In both these cases the relevant energy gap for each of the links is   uniquely  defined either as the gap between the ground state and the first excited state of the system or  just due to the fact that each link is associated with a two-dimensional  Hilbert space and is characterised by a single energy gap to begin with.  This fact allows, in particular, to apply SDRG for the construction of a full basis of (approximate) eigenstates in e.g.\ generalized quantum Ising model of Ref.~\cite{Pekker14}.

\begin{figure}
\includegraphics[width=210pt]{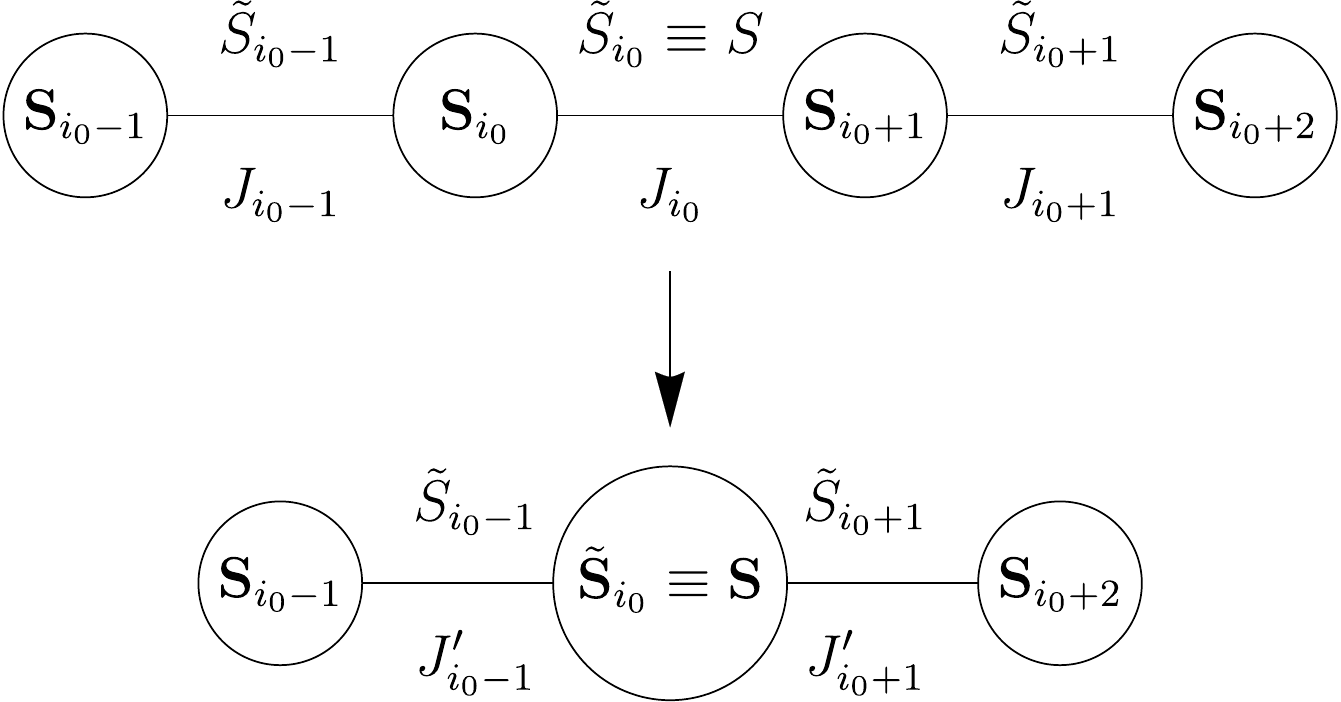}
\caption{Transformation of spins under SDRG. $i_0$ denotes the strongest links to be removed.  After the SDRG transformation the spins $\tilde{S}_{i_0-1}$ and $\tilde{S}_{i_0+1}$ are regenerated iff they are no longer consistent with the rules of angular momentum addition. 
 }
\label{Fig:RGSpins}
\end{figure}

The dependence of the definition of the strongest-coupled subsystem of the spins $\tilde{S}_i$ makes it difficult to generate the full set of eigenstates by the SDRG procedure\footnote{Of course, this is still possible in {\it short} systems at sufficiently strong disorder where the fluctuations in the energy gaps from link to link are dominated by the fluctuations of couplings with the spin-depended factor in Eq. (\ref{App:Eq:Eis})  providing only a numerical coefficient. }. We do not attempt to solve this problem here. Instead, in our numerical analysis we resort to the probabilistic sampling of the SDRG tree states in the middle of the many-body band. To this end we argue that in the infinite temperature ensemble the probability for a couple of spins   to have the total spin $S$ is dominated by the entropic factor $(2S+1)$.
Therefore, starting from the initial SDRG state with $S_i$ representing the elementary spins for each link in the system  we  generate randomly $\tilde{S_i}$ with probability $p(\tilde{S}_i)\propto (2\tilde{S}_i+1)$. We then use Eqs. (\ref{App:Eq:Eis}) and (\ref{App:Eq:Delta}) find the gaps associated  to the links and identify the link $i_0$ to be removed by the SDRG. The link $i_0$ is removed and the pair of spins ${\bf S}_{i_0}$ and ${\bf S}_{i_0+1}$ is replaced by a new spin $\tilde{\bf{S}}_{i_0}$ (see Fig. \ref{Fig:RGSpins}). Before the removal the  spins ${\bf S}_{i_0}$ and ${\bf S}_{i_0+1}$ had spins ${\bf S}_{i_0-1}$ and ${\bf S}_{i_0+2}$ as their left and right nearest neighbors respectively.   The corresponding links, $i_0-1$ and $i_0+1$, had the spins $\tilde{S}_{i_{\rm max }-1}$ and $\tilde{S}_{i_{\rm max }+1}$ associated to them. After the removal of the link $i_0$ the spins  ${\bf S}_{i_0-1}$ and  ${\bf S}_{i_0+2}$ become the neighbors of the spin  $\tilde{\bf{S}}_{i_0}$. We check at this point if the values 
of $\tilde{S}_{i_{0}-1}$ and $\tilde{S}_{i_{0 }+1}$ are still consistent with the rules of the angular momentum addition for the new spin configuration. If this case we keep them as the spin values associated to the newly created links (see Fig. \ref{Fig:RGSpins}). Otherwise,  the newly created links receive  new randomly generated values of the associated spins.

An important ingredient of the SDRG procedure is the renormalization of spin--spin couplings. In the zeroth order of perturbation theory the two strongly-interacting spins ${\bf S}_{i_{0}}$ and ${\bf S}_{i_{0 }+1}$ are treated as  decoupled from the rest of the system.  To establish the coupling of the newly created block spin $\tilde{\bf S}_{i_0}$ to the outside world one needs to take into account the  higher order terms of perturbation theory in the interactions $J_{i_0-1}$ and $J_{i_0+1}$ (see Fig. \ref{Fig:RGSpins}). 
Specifically, we consider the $4$-spin Hamiltonian 
\begin{equation}
H=J_{i_0-1} {\bf S}_{i_0-1} \cdot {\bf S}_1+J_{i_0} {\bf S}_{i_0} \cdot {\bf S}_{i_0+1}+J_{i_0+1}{\bf S}_{i_0+1} \cdot {\bf S}_{i_0+2}.
\label{App:Eq:H4}
\end{equation}
and integrate out fast fluctuations of  ${\bf S}_{i_{0}}-{\bf S}_{i_{0 }+1}$. In a generic case the first order treatment suffices and we end up with the effective Hamiltonian after an SDRG step 
\begin{eqnarray}
H^{(1)}_{\rm eff}&=&J^\prime_{i_0- 1} {\bf S}_{i_0-1}\cdot {\bf S}+J^\prime_{i_0+ 1}{\bf S}\cdot {\bf S}_{i_0+2},\\
J^\prime_{i_0\pm 1}&=&\frac{J_{i_0-1}(|{\bf S}|^2\mp v)}{2{|\bf S}|^2}.
\label{App:Eq:H1}
\end{eqnarray}
Here and below to simplify our notations we denote the spin $\tilde{{\bf S}}_{i_0}$ simply by ${\bf S}$;  the shorthand notation $v$ stands for 
\begin{equation}
v=|{\bf S}_{i_0}|^2-|{\bf S}_{i_0+1}|^2.
\end{equation}

On going over from the Hamiltonian (\ref{App:Eq:H4}) to the Hamiltonian (\ref{App:Eq:H1}) we simply  project the spin vectors ${\bf S}_{i_0}$ and ${\bf S}_{i_0+1}$ on the direction of the (approximately) conserved spin ${\bf S}$. Such an approximation is not sufficient however if the spins  ${\bf S}_{i_0}$,  ${\bf S}_{i_0+1}$ and ${\bf S}$ form a ``quantum pythagorean triangle'', i.e.\ satisfy one of the two conditions
\begin{equation}
|{\bf S}|^2\pm(|{\bf S }_{i_0}|^2- |{\bf S }_{i_0+1}|^2)=0.
\label{App:Eq:pyth}
\end{equation}
One of the couplings $J^\prime_{i_0\pm 1}$ turns then to zero cutting the chain into two independent pieces and the perturbation theory should be developed further. 

A particular case of Eq. (\ref{App:Eq:pyth}) is the singlet formation: $S=0$, $S_{i_0}=S_{i_0+1}$. In that situatiopn the second order perturbative Hamiltonian takes the form \cite{Westerberg1997}
\begin{eqnarray}
H^{(2)}_{\rm eff}&=&\tilde{J} {\bf S}_{i_0-1}\cdot {\bf S}_{i_0+2},\\
\tilde{J}&=&\frac{2 J_{i_0-1}J_{i_0+1}}{3 J_{i_0}}|{\bf S}_{i_0-1}|^2.
\label{App:Eq:H21}
\end{eqnarray}
Note that the singlet formation is the only instance  of Eq.~(\ref{App:Eq:pyth}) relevant in the context of SDRG near the ground state.  

In a more general case of $S\neq 0$ straightforward but lengthy algebra leads to 
\begin{multline}
H_{\rm eff}^{(2)}=\frac{g_{\alpha\beta}}{4J_0}
\left(J_{i_0+2} {\bf S}_{i_0+2}^\alpha-J_{i_0-1}{\bf S}_{i_0-1}^\alpha \right)\\\times\left(J_{i_0+2} {\bf S}_{i_0+2}^\beta-J_{i_0-1}{\bf S}_{i_0-1}^\beta \right)
\label{App:Eq:H22}
\end{multline}
where \footnote{Dealing with the special case of $S=1/2$ in Eq. (\ref{App:Eq:g}) one must take into account the fact that for spin $1/2$ operators ${\bf S}^{\alpha}{\bf S}^\beta=\delta_{\alpha\beta}/4+i\epsilon_{\alpha\beta\gamma}{\bf S}^{\gamma}/2$. The singular denominator in Eq. (\ref{App:Eq:g}) is then cancelled.  }
\begin{widetext}
\begin{multline}
g_{\alpha\beta}=
\frac{1}{4|{\bf S}|^2-3}\left\{\left[2u+|{\bf S}|^2-\frac{3v^2}{|{\bf S}|^2}\right]\delta_{\alpha\beta}-i
\left[-4u-3+2|{\bf S}|^2+\frac{v^2}{|{\bf S}|^4}(3+2 |{\bf S}|^2)\right]\epsilon_{\alpha\beta\gamma}{\bf S}^\gamma \right.\\\left.+
\left[-6u-3+|{\bf S}|^2+\frac{v^2}{|{\bf S}|^4}(5 |{\bf S}|^2+3)\right]\frac{{\bf S}^\alpha {\bf S}^\beta}{|{\bf S}|^2}\right\}.
\label{App:Eq:g}
\end{multline}
\end{widetext}
Here, we denote by $\epsilon_{\alpha\beta\gamma}$ the Levi--Civita  tensor and
\begin{equation}
u=|{\bf S}_{i_0-1}|^2+|{\bf S}_{i_0+2}|^2.
\end{equation}
Note that while the explicit expression for $g_{\alpha\beta}$ is complicated, its tensor structure is fully determined by the $SU(2)$ symmetry.  

Equation (\ref{App:Eq:H22}) shows that the form of the Hamiltonian (\ref{App:Eq:Ham}) is not preserved  under the SDRG transformation if the second order terms are taken into account. We argue however that, while being extremely important for the SDRG flow in the case of antiferromagnetic spin chains near the ground state, the second order renormalizations    play minor role for the physics at infinite temperature. The reason for this is the growth of spins under the SDRG transformation and the fact that in the set of all triples $(S,\, S_{i_0},\, S_{i_0+1})$ the ``pythagorean'' ones have measure zero.

Correspondingly, instead of treating Eq. (\ref{App:Eq:H22})   in its full form, we apply to it several (generically uncontrolled) approximations.  First, we focus on the coupling of spin ${\bf S}$ to that of the spins ${S}_{i_0-1}$ and  ${S}_{i_0+2}$ for which the corresponding first-order coupling in Eq. (\ref{App:Eq:H1}) vanishes. For example, in the case of the plus sign in Eq. (\ref{App:Eq:pyth}) [leading to vanishing of $J^\prime_{i_0-1}$ in Eq. (\ref{App:Eq:H1})]   we make a replacement
\begin{equation}
H^{(2)}_{\rm eff}\rightarrow \frac{J_{i_0-1}^2g_{\alpha\beta}}{4J_0}
{\bf S}_{i_0-1}^\alpha {\bf S}_{i_0-1}^\beta
\label{App:Eq:H22Simple}
\end{equation}
Explicitly, using Eq. (\ref{App:Eq:g}) (and taking into account that $v=S^2$ in the present case) we find 
\begin{multline}
H^{(2)}_{\rm eff}\rightarrow \frac{J_{i_0-1}^2\left(u-|{\bf S}|^2\right)}{2J_{i_0}(4|{\bf S}|^2-3)}\\\times \left[|{\bf S}_{i_0-1}|^2
-2({\bf S}\cdot {\bf S}_{i_0-1})
-\frac{3({\bf S}\cdot {\bf S}_{i_0-1})^2}{|{\bf S}|^2}\right].
\label{App:Eq:App:Eq:H22Simple1}
\end{multline}
Finally, we use the {\it expected} value of the spin $\tilde{S}_{i_0-1}$  (see Fig. \ref{Fig:RGSpins}) to estimate the last term in Eq. (\ref{App:Eq:App:Eq:H22Simple1}) in a kind of mean-field approximstion according to
\begin{equation}
({\bf S} \cdot {\bf S}_{i_0-1})^2\rightarrow \frac12{\bf S} \cdot {\bf S}_{i_0-1} \left[
|\tilde{\bf S}_{i_0-1}|^2-|{\bf S }|^2-|{\bf S }_{i_0-1}|^2
\right].
\end{equation}

After the manipulations outlined above the Hamiltonian $H^{(2)}_{\rm eff}$ reduces back to the Heisenberg model expected by SDRG. We stress that, despite uncontrolled,  the approximations we employ are expected to produce correct order-of-magnitude estimate for the coupling of spins ${\bf S}_{i_0-1}$ and ${\bf S}$ (that vanished in the first order perturbation theory). This should be enough to capture the physics at infinite temperature because the situations when the second order perturbation theory has to be applied are rare. 

\begin{figure}[t]
\includegraphics[width=200pt]{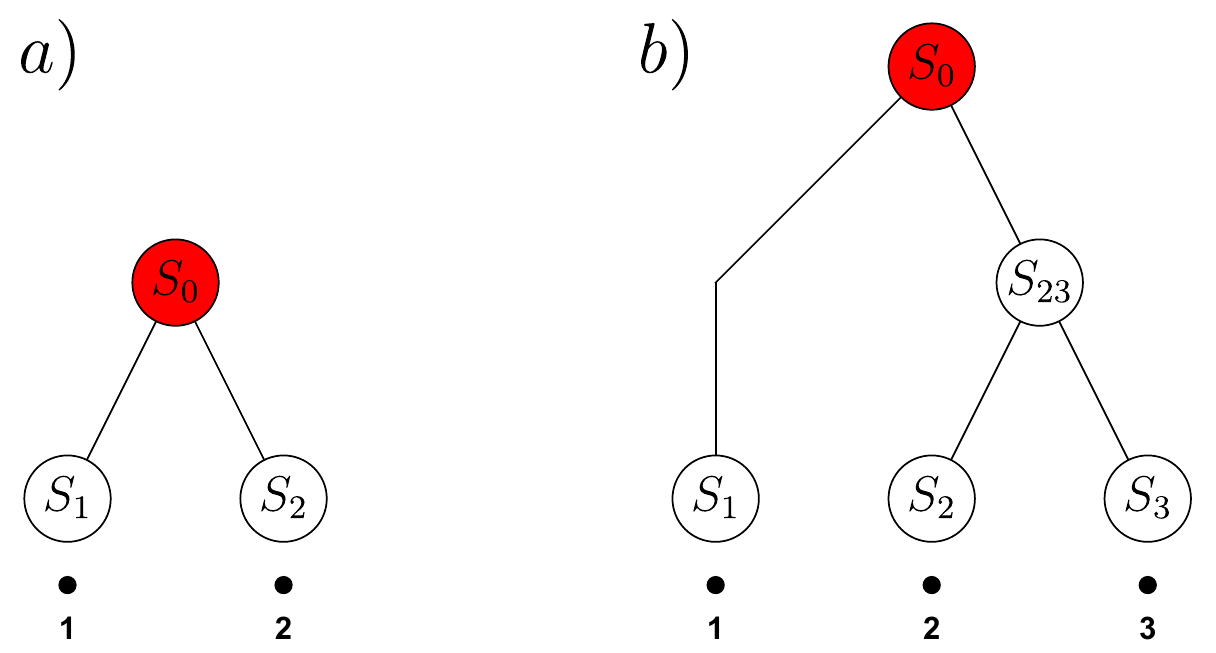}
\caption{ SDRG trees representing states for a system of 2 (a) and 3 (b) spins. Explicit wavefunctions corresponding to these trees are give by Eqs. (\ref{Eq:state2}) and (\ref{Eq:state3}). 
 }
\label{Fig:smallTrees}
\end{figure}

Before closing this Section let us stress once again that in the present work we are primarily interested in the properties of  wavefunctions generated by the SDRG: SDRG tree states. Each  tree generated by SDRG  describes the way the elementary spins in the system fuse to organize an approximate eigenstate of the Hamiltonian (or rather an $SU(2)$-multiplet thereof).   Given an SDRG tree  one can use the  Clebsch-Gordan coefficients  to write down the corresponding wavefunction  in terms of the elementary spin degrees of freedom. We illustrate this process for the  two SDRG trees  shown in Fig. \ref{Fig:smallTrees}. 

In a system of two spins ${\bf S}_1$ and ${\bf S}_{2}$
  a wavefunction with total spin $S_0$ and the $z$-projection of the total spin $M_0$,  $-S_0\leq M_0\leq S_0$, corresponding to the tree shown in Fig.~\ref{Fig:smallTrees}a   reads
\begin{multline}
|S_0, M_0\rangle=\\\sum_{M_1, M_2}C(S_0, M_0; S_1, M_1, S_2, M_2)|S_1, M_1\rangle |S_2, M_2\rangle. 
\label{Eq:state2}
\end{multline}
Here, $|S_i, M_i\rangle$, $i=1, 2$,  is the states of the spin ${\bf S}_i$ with the $z$-axes projection $M_i$. 

In a similar manner for a system of free spins ${\bf S}_1$,  ${\bf S}_{2}$ and ${\bf S}_3$  the wave function corresponding to the tree of Fig.~\ref{Fig:smallTrees}b reads
\begin{widetext}
\begin{multline}
|S_0, M_0\rangle=\sum_{M_1, M_{23}}C(S_0, M_0; S_1, M_1, S_{23}, M_{23})|S_1, M_1\rangle |S_{23}, M_{23}\rangle\\
=\sum_{M_1, M_2, M_3,  M_{23}}C(S_0, M_0; S_1, M_1, S_{23}, M_{23})
C(S_{23}, M_{23}; S_2, M_2, S_{3}, M_{3})
|S_1, M_1\rangle |S_{2}, M_{2}\rangle |S_{3}, M_{3}\rangle
\label{Eq:state3}
\end{multline}
\end{widetext}
Here, by $ |S_{23}, M_{23}\rangle$ we denote the state of the subsystem  made of spins ${\bf S}_2$ and ${\bf S}_3$ with the total spin $S_{23}$ and the spin projection $M_{23}$.  On going from the first to second line in Eq. (\ref{Eq:state3}) we have reexpressed $|S_{23}, M_{23}\rangle$ in terms of 
$ |S_{2}, M_{2}\rangle$ and  $|S_{3}, M_{3}\rangle$.

\section{Entanglement entropy of tree states}
\label{App:Entanglement}

In this Appendix we discuss the entanglement properties of the tree states. 

Lets us consider a single tree state $\left|\Psi\right\rangle$ in a system of $L$ spins $1/2$, see Fig.~\ref{Fig:treeCut}. 
We are interested in the entanglement entropy
\begin{equation}
S_{\rm ent}(L/2)=-\mathrm{Tr}(\rho_{L/2}\log_2\rho_{L/2})
\end{equation}
where $\rho_{L/2}$  stands for the density matrix of e.g.\ the left half of the system. 

To estimate $S_{\rm ent}(L/2)$ we observe that the Schmidt cut in the middle of the chain naturally gives rise to a cut of the tree representing the state into a ``forest'' and a decomposition of the chain into a collection of clusters in the manner exemplified  in Fig. \ref{Fig:treeCut}.  We denote by ${\cal L}_i$ (${\cal R}_i$) the clusters to the left (right) from the cut. It can be readily seen that with the whole system in the state $\left|\Psi\right\rangle$ the quantum state of each of the clusters described above lies in the multiplet specified by the sub-tree build above that cluster. In particular, all the clusters have well defined total spin. The only degree of freedom for each cluster that is not locked by the state  $\left|\Psi\right\rangle$ is the projection of its total spin. It follows then that the rank of the density matrix $\rho_{L/2}$ is limited by
\begin{equation}
{\rm rank}\left(\rho_{L/2}\right)\leq \prod_{{\cal L}_i}(2S_{{\cal L}_i}+1)
\label{Eq:rank}
\end{equation}
where the product runs over all the clusters to the left of the cut and $S_{{\cal L}_i}$ are corresponding total spins. 

\begin{figure}[t]
\includegraphics[width=240pt]{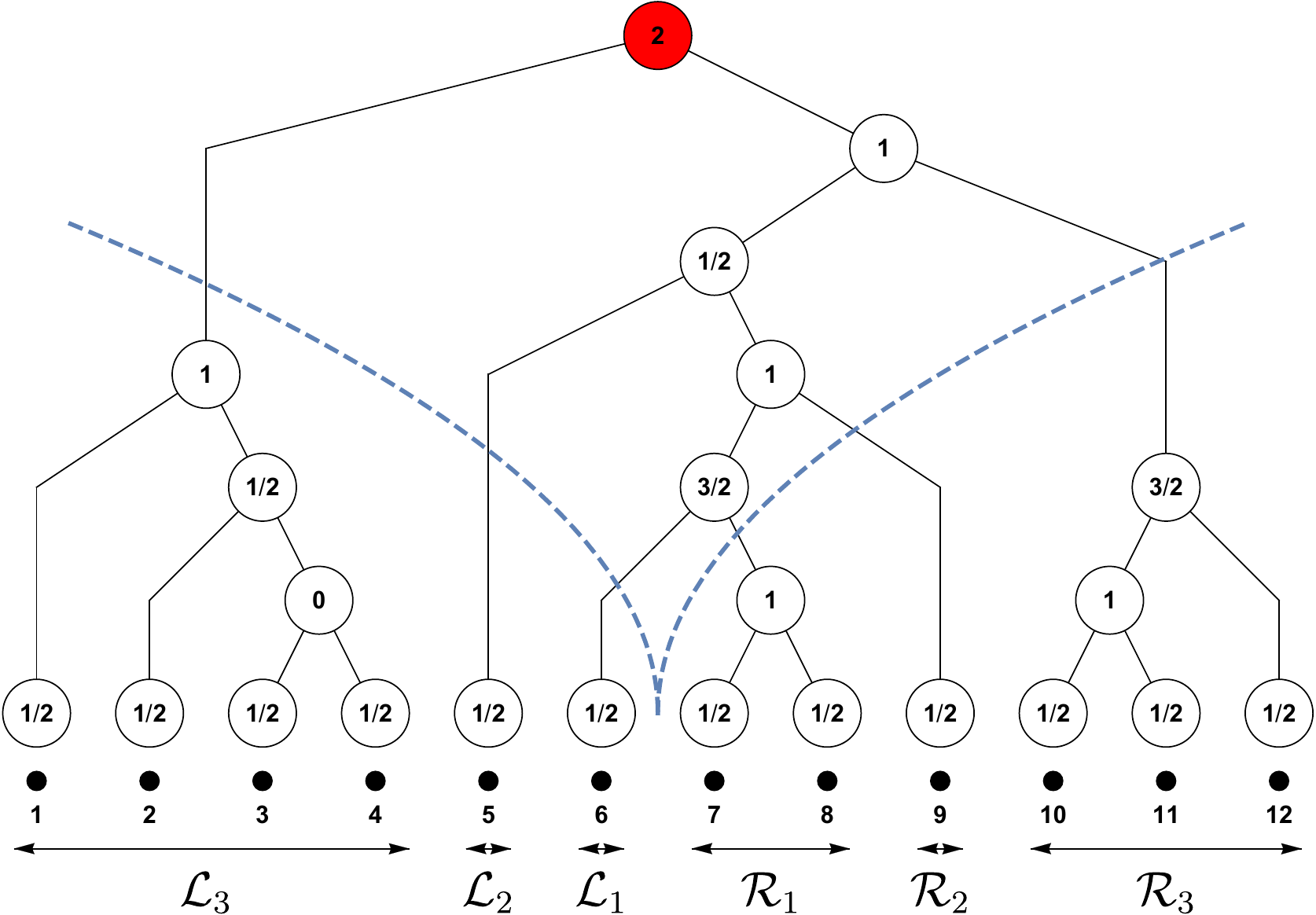}
\caption{ Tree state generated by SDRG and its entanglement properties.  A Schmidt cut at the middle of the system gives rise to a cut of the  tree into a ``forest'' and  prescribes a view of the chain as a collection of clusters lying to the left (${\cal L}_i$) and to the right (${\cal R}_i$) of the cut. With the full chain in the quantum state described by the tree each cluster has the projection of the total momentum as the only degree of freedom. 
 }
\label{Fig:treeCut}
\end{figure}

Each of the spins $S_{{\cal L}_i}$ is limited by  $L/2$ while the number of clusters can mot exceed the depth $d$ of the tree. Correspondingly, the entanglement entropy of the tree state $\left|\Psi\right\rangle$ satisfies
\begin{equation}
S_{\rm ent}(L/2)\leq \log_2 {\rm rank}\left(\rho_{L/2}\right) \leq d \log_2 L.
\label{Eq:S1}
\end{equation}
In the case of a logarithmic tree, $d\sim \log_2 L$,  Eq. (\ref{Eq:S1}) implies the estimate
\begin{equation}
S_{\rm ent}(L/2)< c \log^2_2 L
\label{App:Eq:SAbove}
\end{equation}
with some numerical constant $c$ of order $1$ that depends on the statistical properties of the tree. 
This proves the upper bound for the entanglement stated in Eq. (\ref{Eq:SUp}) in the main text.
 
From the consideration above we see that the smallest value of entanglement is to be expected 
when the Schmidt cut at the middle of the  chain cuts the tree  into just two subtrees [so that there is only one left and one right cluster (${\cal L}_1$ and ${\cal R}_1$) in Fig. \ref{Fig:treeCut}].  
Let us denote the the spins of the left and right cluster by $S_{\cal L}$ and $S_{\cal R}$ respectively. The density matrix $\rho \equiv \rho_{L/2}$ depends  on the total spin $S$ and its projection $M$  in the state $\left|\Psi\right\rangle$. It  is of dimension $(2 S_{\cal L}+1)$  and can be written explicitly in terms of Clebsch-Gordan coefficients $C$
\begin{multline}
\rho_{M_{\cal L}M_{\cal L}^\prime}=\sum_{M_{\cal R}}C(S, M; S_{\cal L}, M_{\cal L}, S_{\cal R}, M_{\cal R})\\\times C(S, M; S_{\cal L}, M^\prime_{\cal L}, S_{\cal R}, M_{\cal R})\, \quad  -S_{\cal L} \leq M_{\cal L}, M^\prime_{\cal L}\leq S_{\cal L}
\end{multline}
The conservation of the projection of the angular momentum forces then the density matrix to be diagonal
\begin{equation}
\rho_{M_{\cal L}M^\prime_{\cal L}}=\rho_{M_{\cal L}}\delta_{M_{\cal L}M^\prime_{\cal L}}
\end{equation}
where 
\begin{equation}
\rho_{M_{\cal L}}=C^2(S, M; S_{\cal L}, M_{\cal L}, S_{\cal R}, M-M_{\cal L}).
\label{App:Eq:rho1}
\end{equation}

Particularly simple case is that of  $S=0$ (which implies $S_{\cal L}=S_{\cal R}$ and $M=0$)  where Eq. (\ref{App:Eq:rho1}) reduces to
\begin{equation}
\rho_{M_{\cal L}}=\frac{1}{2S_{\cal L}+1}
\end{equation}
and gives the entanglement entropy 
\begin{equation}
S_{\rm ent}(L/2)=\log_2 \left(2S_{\cal L}+1\right).
\label{App:Eq:S1}
\end{equation}

For a typical tree state $S_{\cal L}\propto \sqrt{L}$ and the entanglement entropy 
\begin{equation}
S_{\rm ent}(L/2)\propto \frac12 \log_2 L
\end{equation}
in agreement with Ref. \cite{Protopopov2017}.

While we have no proof of the logarithmic scaling of entanglement for arbitrary values of $S$, $M$ 
$S_{\cal L}$ and $S_{\cal R}$ in Eq. (\ref{App:Eq:rho1}) (and this scaling certainly does {\it not} hold in some specific cases, e.g.\ $S=S_{\cal L}+S_{\cal R}$, $M=S$) we expect that the lower bound on entanglement
\begin{equation}
S_{\rm ent}(L/2)\gtrsim c \log_2 L
\end{equation}
stated in Eq. (\ref{Eq:SUp}) in the main text remains correct for the typical tree states. 

The upper bound on the entanglement entropy, Eq. (\ref{App:Eq:SAbove})  can be easily generalized to the case when the state of interest is not a single tree state but a superposition of a finite number thereof, $n_{\rm T}$.   The rank of the density matrix in this case is limited by [cf. Eq. (\ref{Eq:rank}) ]
\begin{equation}
{\rm rank}\left( \rho_{L/2}\right)\leq n_{\rm T} L^d 
\end{equation}
and the entanglement entropy satisfies
\begin{equation}
S_{\rm ent}{(L/2)}<  c \log_2^2 L+\log_2 n_{\rm T}
\end{equation}
We conlude that the entanglement entropy grows logarithmically with the number of tree states involved and of the order of $2^{L/2}$ of them are required to recover the volume-law scaling of ergodic eigenstates.

\section{Searching for resonances}
\label{App:ResonanceSearch}
In this appendix we briefly review our numerical procedure for searching resonances. 

Let us consider a tree state $\left|\Psi^{0}_{\rm RG}\right\rangle$ generated by the~SDRG. Fixing the tree geometry but allowing the values of the block spins in the non-leaf nodes of the tree to take arbitrary values consistent with the rules of the angular momentum addition provides us with the basis in the Hilbert space. In Sec. \ref{Sec:ShortScales_2}  such a basis was denoted by  $\left|\Psi^{a}_{\rm RG}\right\rangle$ (with $a=1\,,\ldots D_{S_0, L}$). We are interested in the matrix elements of the Hamiltonian, $H_{0a}$  between the original RG state $\left|\Psi^{0}_{\rm RG}\right\rangle$ and other members of the basis. Of particular importance for us are the  resonant situations when $|H_{0a}|> |H_{aa}-H_{00}|$.

The dimension of the Hilbert space $D_{S_0, L}$ scales  exponentially with the length $L$.  Fortunately,  most of the matrix elements of the Hamiltonian are, in fact, identically zero due to the $SU(2)$ symmetry. For an arbitrary pair of spins $i$ and $j$  the operator ${\bf S}_i\cdot{\bf S}_j$  acting on $\left|\Psi^{0}_{\rm RG}\right\rangle$   can only change those block spins that lie on the path in the tree connecting the spins $i$ and $j$. Moreover, the selection rules  analogous to the ones in optics limit possible change in each block spin $S$  to $\Delta S=\pm 1$ or $0$. In addition, $\Delta S=0$ is forbidden in the case of $S=0$. 

Using this selection rules  together with the fact that the Hamiltonian is just a linear combination of operators ${\bf S}_{i}\cdot{\bf S}_{i+1}$ we are able to count and index all the states    $\left|\Psi^{a}_{\rm RG}\right\rangle$ such that  $H_{a0}\neq 0$ (we call them the  neighbors of the state  $\left|\Psi^{0}_{\rm RG}\right\rangle$) without actually generating them. We denote by $K$ the number of available neighbors.

We then start the random search of resonances among the neighbours. To pick a random neighbour we generate a random integer from an interval $[1, K]$ and recompute the corresponding neighbour. We then evaluate the matrix element $H_{0a}$ and the energy difference $H_{aa}-H_{00}$\footnote{An efficient way to evaluate $H_{ab}$ was discussed in Ref.~\cite{Protopopov2017}}. If the resonance condition is met we record the information about the resonant neighbour. The random search runs over some number $n_{\rm samp}$ of neighbors. Depending on the size of the system   $n_{\rm samp}$ can reach values up to~$2\times 10^7$. Given the number $n_{\rm res}$  discovered during the random sampling we estimate the total number of resonant neighbors of the given SDRG state by
\begin{equation}
K_{\rm res}=\frac{n_{\rm res}}{n_{\rm samp}} K.
\end{equation}
Note that for system sizes $L<2000$ the sampling number  $n_{\rm samp}$ actually {\it exceeds} the total number of available neighbours $K$ so that the random sampling could  be replaced by exhaustive search. For larger system sizes   an exhaustive search becomes, however, unfeasible.

We repeat the procedure outlined above for $5000$ of disorder realisations generating for each disorder realization a single SDRG state. In long systems $K_{\rm res}$ (as well as $K$) fluctuates strongly from sample to sample. The $\ln K_{\rm res}$ develops however a well-behaved distribution exemplified   in Fig. {\ref{Fig:ResNumberDistr}}. Therefore, in the main text we characterise the proliferation of resonances in long systems by the typical value of $K_{\rm res}$, $e^{\langle\ln K_{\rm res} \rangle}$, whose dependence on the size of the system was discussed in Sec. \ref{Sec:Resonances_3}. 

\begin{figure}
\includegraphics[width=230pt]{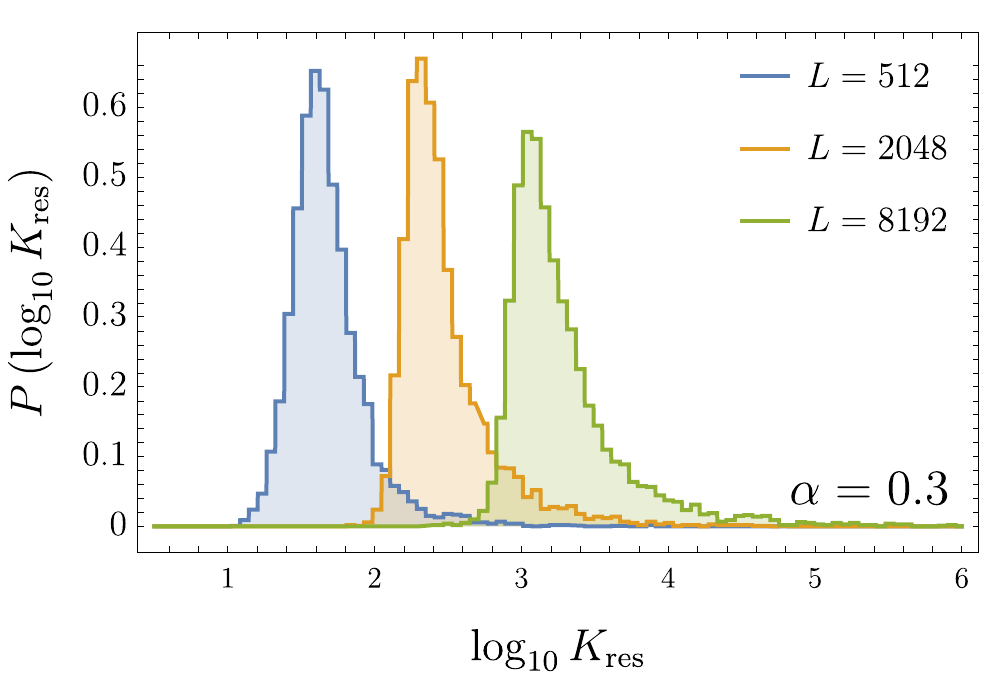}
\caption{The probability distribution $P\left(\log_{10} K_{\rm res}\right)$ at $\alpha=0.3$.  
 }
\label{Fig:ResNumberDistr}
\end{figure}

\section{Statistical properties of SDRG trees}

In this Section we study the properties of \emph{random} SDRG trees. These are obtained by randomly picking spins to fuse together, retaining only their spatial arrangement, ignoring the values of the (bare or renormalized) $J$'s. This simplification allows us to get some analytical results.

\subsection{Distribution of nearest-neighbor graph distances}
\label{App:distance_distribution}
We want to prove the claim (\ref{eq:PL}) in the main text,
\begin{equation}
P(l) = \frac{3}{4}\left(\frac{2}{3}\right)^l,\quad\text{for \( L\to\infty \)},
\end{equation}
where $P(l)$ is the distribution of the random variable $l_{i,i+1}$, namely the graph distance of two neighboring spins in a generic SDRG tree.

To this end, we consider the ensemble of trees constructed by taking a chain of $L$ spins and fusing them all together, two neighbors at a time. After each fusion, the chain effectively shrinks by one site, and the neighbor structure gets updated accordingly.
This is an approximation of the SDRG procedure where we completely neglect the detailed structure of the $ J $ couplings.

More precisely, let a tree be described by a sequence of fusions $ (i_1, \dotsc, i_{L-1}) $, where $ i_k $ means that we are fusing, at the $ k $-th step, the pair $ (i_k, i_k+1) $ (with periodic boundary conditions). In order to emulate the SDRG algorithm, we sample the sequence of fusions uniformly randomly among the $ L! $ possible $ (L-1) $-permutations of $ (1, \dotsc, L) $. This results in a \emph{biased} distribution on the set of all binary trees, with ``taller'' trees being less likely.

Now take a generic pair $ (i,i+1) $ in a given tree, and suppose that their common block-spin descendant was created at the $ (k+1) $-th step of the tree construction. All the fusions taking place after that step are irrelevant for determining $ l_{i,i+1} $, whereas each of the $ k $ previous ones may contribute either 0 or 1 to such distance. In fact, the distance contributed by the $ j $-th fusion is a Bernoulli random variable with success probability $ p_j = \frac{2}{L-j} $, because the distance between $ i $ and $ (i+1) $ only increases if either one of their descendants is picked out of the $ L-j $ possible spins at that step. Moreover, the contributions are uncorrelated since all free indices are sampled with equal probability regardless of the previous history of the tree construction. 

Therefore we have
\begin{equation}
l_{i,i+1}=l^{(k)} = 2+x_1+x_2+...+x_{k}
\label{eq:lrand}
\end{equation}
where
\begin{equation}
x_j=
\begin{cases}
1\quad \mbox{ with probability } &p_j=\frac{2}{L-j}\\
0\quad \mbox{ \quad\ " \qquad " }&1-p_j\\
\end{cases}
\label{eq:xjrand}
\end{equation}
and the $(k)$ superscript serves as a reminder that our random variable is now being conditioned on $k$.

Let us compute the cumulant generating function for $ l^{(k)}-2 $:
\begin{equation}
\langle e^{-s(l^{(k)}-2)}\rangle=\prod_{j=1}^k(1-p_j+p_j e^{-s}),
\end{equation}
the logarithm of which is
\begin{equation}
\ln\langle e^{-s(l^{(k)}-2)}\rangle=\sum_{j=1}^k\ln\left(1+\frac{1}{L}\frac{2}{1-j/L}(e^{-s}-1)\right).
\end{equation}
By defining $ x = j/L $, $  \alpha =k/L $, and taking $L\to\infty$, we have
\begin{eqnarray}
\ln\langle e^{-s(l^{(k)}-2)}\rangle&\sim&\int_0^\alpha \mathrm{d}x\frac{2}{1-x}(e^{-s}-1)\nonumber\\
&=&2(1-e^{-s})\ln(1-\alpha)
\end{eqnarray}
up to $ O(1/L) $ terms.

Now notice that $ (k+1) $, in our ensemble, is uniformly distributed between 1 and $ L-1 $, as it corresponds to the position of index $ i $ in the tuple $ (i_1, \dotsc, i_{L-1}) $.
We can then get rid of the $k$-conditioning by averaging over $\alpha \in [0,1]$. This gives
\begin{eqnarray}
\langle e^{-s(l-2)}\rangle&=&\int_0^1\mathrm{d}\alpha (1-\alpha)^{2(1-e^{-s})}\nonumber\\
&=&\frac{1}{3-2e^{-s}},
\end{eqnarray}
and by expanding the denominator in a geometric series, we get
\begin{equation}
\langle e^{-sl}\rangle=\frac{1}{3}\left[e^{-2s}+\frac{2}{3}e^{-3s}+\left(\frac{2}{3}\right)^2e^{-4s}+\dotsc\right].
\end{equation}
Inverting the Laplace transform results in Eq.\ (\ref{eq:PL}).

\subsection{Size of the block spins}
\label{App:size_block_spins}
We now set out to determine the average size of the support of a randomly chosen block-spin operator for a random SDRG tree state. This amounts to estimating the number of leaves which connect to a node picked uniformly randomly from the set of non-leaf nodes in a generic fusion tree.

To this end, it is convenient to introduce an alternative (but equivalent) construction for our random ensemble. Consider a single node, and start by attaching two children nodes to it, one to the left and one to the right. We can see this as a ``splitting'' step for the original node. Now pick with equal probability either one of the resulting leaves and perform the same kind of splitting. Iterate the procedure for a total of $ (L-1) $ times, such that the final number of leaves is $L$. The leaves are spacially ordered by the order relation induced in an obvious way by the distinction of left- and right-children. In this way we obtain a binary tree whose geometry is compatible with an SDRG tree. We can call this the ``fission tree'' ensemble.

We are now going to prove by induction that the fission tree and fusion tree ensembles are equivalent \footnote{The following proof was provided by user Misha Lavrov as part of a reply to a question on the \href{https://math.stackexchange.com}{math.stackexchange.com} website. The authors would like to acknowledge the contribution and thank StackExchange for providing a platform to discuss such topics.}.

Suppose that the above claim holds after the $(k-1)$-th splitting, that is to say, for the ensembles of $k$-leaved fission and fusion trees.
Now, when constructing a \emph{fusion} tree on $(k+1)$ leaves, after the first fusion we are left with an effective $k$-leaved tree. In order to prove the claim it is then enough to show that the first fusion does not spoil the ensemble equivalence.
By definition of the fusion tree ensemble, it is the case that each one of the initial $(k+1)$ leaf pairs has the same probability of being fused at the first step, which means that every one of the $k$ effective leaves after the first step has the same likelihood of being the one resulting from the fusion. Therefore, upon reversing the ``time direction'' we see that if we allow all the $k$ leaves to split with the same probability, both fission and fusion trees on $(k+1)$ leaves are sampled with the same distribution, and the inductive step is completed. It also holds trivially that the two ensembles coincide when $k = 1$, providing the basis of the induction.

In light of this, it is possible to assign to each node of a tree the step at which it was split. For instance, the root will always be labeled by 1, and the maximum label will be $ L-1 $ (note that, similarly to the case of the ``fusion labeling'' in Appendix~\ref{App:distance_distribution}, this labeling is not uniquely defined).
Now fix $k\in\{1,\dotsc,L-1\} $ and consider the node labeled by $k$. Introduce the variable $t$ to measure the number of fissions occurring after the $k$-th one, $t \in \{0, \dotsc, L-k-1\}$, and call $N(t)$ the total number of leaves which affect the state of the initial node at ``time'' $t$. Since every fission can only increment $N$ by 1 at time $(t+1)$ if one of the $N(t)$ leaves is picked for the fission, we have the stochastic recursion equation
\begin{equation}
N^{(k)}(t+1) = N^{(k)}(t) + B[p^{(k)}(t)],
\end{equation}
where $B[p]$ is a Bernoulli variable with success probability $p$, and $p^{(k)}(t) = \frac{N(t)}{k+t+1}$. The initial condition must be set to $N^{(k)}(0) = 2$.

This equation is hard to treat due to the $N(t)$-dependence hidden inside $p^{k}(t)$, but it is linear, and therefore easily solved in the expectation values:
\begin{equation}
\overline{N^{(k)}(t+1)} = \overline{N^{(k)}(t)}\left(1 + \frac{1}{k+t+1}\right),
\end{equation}
where we used $\overline{B[p]} = p$. By iterating and simplifying the product on the right hand side, and then looking at the final time, we get
\begin{equation}
\overline{N^{(k)}} \equiv \overline{N^{(k)}(L-k-1)} = \frac{2L}{k+1}.
\end{equation}

This is the average number of ancestors of a node that was split at the $k$-th fission step. In order to answer our initial question --- what is the average number of ancestor elementary spins of a random non-leaf node ---, we simply take the average on all possible values of $k$. This yields
\begin{equation}
\overline{N} = \frac{1}{L}\sum_{k=1}^{L-1}\overline{N^{(k)}} = 2\left(\log L + \gamma - 1\right) + O\left(\frac{1}{L}\right),
\end{equation}
showing that the block spins have on average unbounded support in space.


\end{document}